\documentclass{article}

\usepackage{arxiv}

\usepackage[utf8]{inputenc}
\usepackage[T1]{fontenc}
\usepackage{hyperref}
\usepackage{url}
\usepackage{booktabs}
\usepackage{amsfonts}
\usepackage{nicefrac}
\usepackage{microtype}
\usepackage{graphicx}
\usepackage{natbib}
\usepackage{doi}
\usepackage[USenglish]{babel}
\usepackage[intlimits]{amsmath}
\usepackage{bm}
\hypersetup{colorlinks=true, citecolor=blue}
\usepackage{amssymb}
\usepackage{fontawesome}
\usepackage{mathtools}
\usepackage{xcolor}
\usepackage{lmodern}
\usepackage{siunitx}
\usepackage{booktabs}
\usepackage{etoolbox}

\renewrobustcmd{\bfseries}{\fontseries{b}\selectfont}
\renewrobustcmd{\boldmath}{}
\newrobustcmd{\B}{\bfseries}

\title{A Critique of a Variety of “Memory-Based” Process Monitoring Methods}

\date{October 20, 2021}

\author{%
	\href{https://orcid.org/0000-0002-9666-5554}{\includegraphics[scale=0.06]{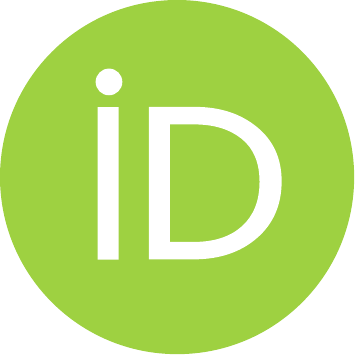}\hspace{1mm}Sven Knoth} \\
	Dep. of Mathematics \& Statistics\\
	Helmut Schmidt University\\
	Hamburg, Germany\\
	\texttt{knoth@hsu-hh.de} \\
	\And 
	\href{https://orcid.org/0000-0002-3618-6746}{\includegraphics[scale=0.06]{orcid.pdf}\hspace{1mm}Nesma A. Saleh} \\
	Dep. of Statistics\\
    Cairo University\\
	Giza, Egypt\\
	\texttt{neasaleh@feps.edu.eg} \\
	\And 
	\href{https://orcid.org/0000-0003-4753-5195}{\includegraphics[scale=0.06]{orcid.pdf}\hspace{1mm}Mahmoud A. Mahmoud} \\
	Dep. of Statistics\\
	Cairo University\\
	Giza, Egypt\\
	\texttt{mamahmou@feps.edu.eg} \\
	\And 
	\href{https://orcid.org/0000-0002-9962-0001}{\includegraphics[scale=0.06]{orcid.pdf}\hspace{1mm}William H. Woodall} \\
	Dep. of Statistics\\
	Virginia Tech\\
	Blacksburg VA, USA\\
	\texttt{bwoodall@vt.edu} \\
	\And
	\href{https://orcid.org/0000-0002-5196-3451}{\includegraphics[scale=0.06]{orcid.pdf}\hspace{1mm}V{\'i}ctor G. Tercero-G{\'o}mez} \\
	School of Engineering \& Sciences\\
	Tecnologico de Monterrey\\
	Monterrey, Nuevo Leon, Mexico\\
	\texttt{victor.tercero@tec.mx} \\
}

\renewcommand{\shorttitle}{The Case against GWMA Charts}

\begin{document}
\maketitle
	
\begin{abstract}
Many extensions and modifications have been made to standard process monitoring methods such as the exponentially weighted moving average (EWMA) chart and the cumulative sum (CUSUM) chart. In addition, new schemes have been proposed based on alternative weighting of past data, usually to put greater emphasis on past data and less weight on current and recent data. In other cases, the output of one process monitoring method, such as the EWMA statistic, is used as the input to another method, such as the CUSUM chart. Often the recursive formula for a control chart statistic is itself used recursively to form a new control chart statistic. We find the use of these ad hoc methods to be unjustified. Statistical performance comparisons justifying the use of these methods have been either flawed by focusing only on zero-state run length metrics or by making comparisons to an unnecessarily weak competitor. 
\end{abstract}


\section{Introduction} \label{sec:intro}

Many extensions and modifications have been made recently to standard process monitoring methods such as the exponentially weighted moving average (EWMA) chart and the cumulative sum (CUSUM) chart. We find that many of these methods add complications with no benefits in terms of statistical performance. In addition, new schemes have been proposed based on alternative weighting of past data, usually to put greater emphasis on past data and less weight on current and recent data. Some of these methods, the homogeneously weighted moving average (HWMA) chart, the progressive mean (PM) chart, and the generally weighted moving average (GWMA) chart, have already been studied by \citet*{Knot:EtAl:2021a} and \citet*{knoth2021case}, and found to be flawed and unnecessary.

We consider in our paper what we refer to as compound charts. These include what are referred to in the literature as mixed or hybrid charts. In these cases the output of one process monitoring method, such as the EWMA or moving average (MA) statistic, is used as the input to another method, such as the CUSUM chart. Increasingly the recursive formula for a control chart statistic is itself used recursively as with the double and triple EWMA charts. We find these methods to be ad hoc, unnecessary, inadequately justified, and often with unreasonable weighting patterns for data where the past data values are given more weight than the present ones. Past comparisons justifying the use of these methods are either based on zero-state run length performance instead of the more realistic steady-state performance, on comparisons made to an unnecessarily weak competitor, or both.

What have been referred to in the literature as “mixed” control charts should not to be confused with the simultaneous use of more than one control chart such as the use of several cumulative sum (CUSUM) charts with different reference values or the use of a Shewhart chart in conjunction with a CUSUM chart, as proposed by \cite{Luca:1982}. Instead, mixed charts involve the use of the control chart statistic of one chart as input into the control chart statistic or rule of another chart. \cite{riaz2011improving}, \cite{Abba:Riaz:Does:2011a}, and \cite{Abba:EtAl:2015} have proposed using runs rules with EWMA and CUSUM charts. These methods fit into our framework.

We aim in this article to provide an extensive review of the literature on the compound control charts. We also show that a proper performance comparison of these charts to the conventional ones shows the added complications provide no benefits. We evaluate the proposed charts in terms of the more realistic steady-state performance and the conditional expected delay (CED). We note that with compound charts, a Markov chain approach is no longer easily derived. We also show that the use of the standard design parameter values of the conventional methods (e.g. observations' weights) provides misleading comparisons and conclusions since these compound approaches change the usual weighting structure on past and current observations. 

Most of the papers we review on compound charts have been published in the last five years. We obviously cannot study the performance of all of these methods, so we chose to study in detail only five of them as illustrations. These are the mixed EWMA-CUSUM chart proposed by \cite{abbas2013mixed}, the use of runs rules with CUSUM and EWMA charts, the double moving average (DMA) chart of \cite{Khoo:Wong:2008}, the double EWMA (DEWMA) chart of \cite{Sham:Amin:Sham:1991} and \cite{Sham:Sham:1992}, and the double PM (DPM) of \cite{Abba:EtAl:2019a}.

The paper is organized as follows.
In Section~\ref{sec:LitRev}, we provide an extensive literature review on the proposed compound charts; namely the recursive EWMA charts, memory-type charts with run rules, MA, PM, HWMA, and mixed control charts.
Afterwards, we provide some basic notation in Section~\ref{sec:basics}.
We re-evaluate each of the mixed EWMA-CUSUM, RR-CUSUM/EWMA, DMA, DEWMA, and DPM methods in Sections~\ref{sec:mec}, \ref{sec:rr_ewma_cusum}, \ref{sec:madmatmaqma}, \ref{sec:dewma}, and \ref{sec:dpm}, respectively,
in terms of their zero- and steady-state performance and their CED behavior. In Section~\ref{sec:conclusions}, we provide our concluding remarks.

\section{Literature Review} \label{sec:LitRev}

\subsection{Recursive use of the EWMA statistic} \label{subsec:ewma}

A recursive use of the EWMA statistic implies switching the EWMA chart statistic formula into a recursive function; by which it keeps calling itself as a function input in a repeated manner. \cite{Sham:Sham:1992} introduced a double EWMA (DEWMA) chart, while \cite{Alev:Chat:Kouk:2021c} introduced a triple EWMA (TEWMA) chart. \cite{haq2013new} introduced a hybrid EWMA (HEWMA) chart which is equivalent in structure and concept to the DEWMA chart. \cite{haq2017new} noted that the variance of the chart statistic derived in \cite{haq2013new} was incorrect, and provided the correct formula. Throughout this section, the DEWMA chart terminology will also be used to refer to the HEWMA chart since the DEWMA and HEWMA charts are equivalent.

By expanding the statistics of these proposed charts into weighted averages, one can easily realize that they are fundamentally flawed in that they give past data values more weight than current values. As discussed by \cite{Lai:1974}, for example, the weight given to a particular data value should not increase as the data value ages. This undesirable characteristic does not adversely affect zero-state run length performance, but it can result in poor steady-state run length performance. Many other compound charts based on recursive use of control chart statistic formulas share this property. Virtually all performance comparisons justifying compound charts are based on the less realistic zero-state performance metrics under the assumption that any process shift occurs immediately as monitoring begins. Giving more weight to past data values than to current data values is clearly not reasonable in process monitoring applications. 

Performance comparisons of the DEWMA or TEWMA chart with the EWMA chart typically use the same smoothing parameter for both methods. A more competitive EWMA chart would be one with a weighting scheme on past data similar to that of the DEWMA or TEWMA chart. Such comparisons then show no performance benefits for the DEWMA or TEWMA chart. \cite{Mahm:Wood:2010} extensively re-evaluated the DEWMA chart in comparison with the EWMA chart in terms of the zero-state performance with adjusted weights and worst-case (inertia-effect) performance. They concluded that the DEWMA chart shows a significantly weak performance in comparison to the EWMA chart because of the higher weights it gives to the older observations. 

Despite its flaws, the DEWMA approach has been implemented in a number of applications. \cite{Zhan:Chen:2005} investigated the DEWMA chart in comparison with the EWMA chart to identify the range of shifts where the former surpasses the latter. They additionally provided some design values for the DEWMA chart. \cite{alkahtani2013robustness} investigated the robustness of the DEWMA chart under non-normality, and \cite{nawaz2021ewma} studied the effect of non-normality on its performance. \cite{raji2018designing} evaluated the use of some robust estimators in monitoring the process location parameters instead of the simple sample mean under known and estimated process parameters. \cite{ahmed2020robust} integrated the DEWMA chart with a generalized least-squares (GLS) algorithm based on order statistics to develop a robust DEWMA chart to monitor non-normal processes.

\cite{perez2017double} used a DEWMA chart to monitor linear drifts/shifts in the quality characteristic of interest. \cite{asif2020hybrid} incorporated measurement error into the model used with the chart, while \cite{noor2020hybrid} took a Bayesian approach with it. \cite{noor2019hewma}, \cite{raza2019hybrid}, \cite{tariq2020improved}, and \cite{haq2021new} incorporated auxiliary information into the DEWMA chart; with \cite{raza2019hybrid} considering two auxiliary variables. Auxiliary variables are those that are highly correlated with the variable of interest. It is assumed that the parameters of the distribution of the auxiliary variables are both known and cannot change over time. In survey sampling literature, such variables are used to increase the efficiency of the estimators of the population parameters. The required assumptions are unlikely to hold in process monitoring applications, as discussed by \cite{Saleh:EtAl:2021a}.

\cite{khoo2010monitoring} proposed a Max-DEWMA chart which is based on the maximum of two DEWMA statistics to detect simultaneously changes in the process mean and variance. \cite{javaid2020new} evaluated it in the presence of auxiliary information, while \cite{javaid2021maximum} evaluated it in the presence of both measurement errors and auxiliary information. Further, for simultaneous monitoring of the process mean and dispersion, \cite{teh2011sum} proposed a sum of squares DEWMA (SS-DEWMA) chart. In the SS-charts, two chart statistics are used; one for the process mean and one for the process variance, and the final chart statistic by which the status of the process is determined is based on the sum of squares of both statistics. \cite{ali2017new} and \cite{tariq2019design} developed a DEWMA chart for monitoring only the process variance, while \cite{aslam2019nonparametric} developed a proportion-DEWMA chart to monitor the process variance under non-normal or unknown (nonparametric) distributed processes.

\cite{azam2015designing} and \cite{adeoti2018new} added a repetitive sampling feature to the DEWMA chart. A repetitive sampling feature allows for taking additional samples at a given time point when the results of the drawn sample are indecisive. \cite{shafiq2018note} corrected and recalculated some designs provided incorrectly in \citeauthor{azam2015designing}'s (\citeyear{azam2015designing}) article. \cite{Noor-ul-amin2020hybrid} evaluated the DEWMA chart under different ranked set sampling (RSS) schemes; which are classical RSS, Extreme RSS (ERSS), Median RSS (MRSS), and Quartile RSS (QRSS). The RSS is a sampling scheme that depends on ranking the collected items when the actual measurements are difficult to make
\citep{mcintyre1952method}.
\cite{Riaz:Abba:2016}
and \cite{Raza:EtAl:2020a}
developed a nonparametric DEWMA chart for monitoring the process location parameter, and \cite{Shaf:EtAl:2020} integrated the repetitive sampling technique with it.
\cite{Chan:EtAl:2021} developed a distribution-free DEWMA chart based on the Lepage statistic to monitor simultaneously the process location and scale parameters. 

On the attribute variable side,
\cite{Zhan:EtAl:2003} evaluated the DEWMA chart under Poisson processes.\cite{Asla:EtAl:2016} applied the chart on monitoring processes that use attribute and variable inspection at the same time. \cite{Asla:Khan:Jun:2018} and \cite{Alev:Kouk:2019a} developed a DEWMA chart to monitor the parameters of the Conway-Maxwell-Poisson (COM) distribution. This distribution is used to represent under-dispersed and over-dispersed count data. \cite{Alev:Kouk:2020e} and \cite{Alev:Kouk:2021b} used a DEWMA chart to monitor zero-inflated Poisson (ZIP) and zero-inflated binomial (ZIB) distributed processes, respectively.   

Other underlying process distributions were also considered. For example, \cite{Alev:Kouk:2020f}, \cite{Raza:EtAl:2020b}, and \cite{Adeo:2020} used a DEWMA chart to monitor time between events (TBE) (gamma-distributed processes), Weibull data, and exponentially distributed processes, respectively. \cite{Bizu:Wang:2019}
proposed a likelihood ratio based DEWMA chart to monitor the shape parameter of the inflated Pareto process.

Further extensions and applications were conducted by
\cite{Abde:EtAl:2016} who used the DEWMA chart for profile monitoring, and \cite{Alka:Scha:2012} and \cite{Kuva:2020}
who extended the doubling concept to the multivariate EWMA chart.

As for the TEWMA chart, \cite{Alev:Chat:Kouk:2021f, Alev:Chat:Kouk:2021i} developed a nonparametric version of the chart, while \cite{Lets:EtAl:2021} did the same but with adding an improved modified FIR (IMFIR) feature to the chart to monitor changes in process location.\cite{Alev:Chat:Kouk:2021j}
proposed the use of a TEWMA chart to monitor the TBE (gamma-distributed processes).\cite{Chat:Kouk:Lapp:2021} used a TEWMA chart for monitoring process dispersion.

\cite{Ali:EtAl:2021b} developed a conditional expected value (CEV) hybrid DEWMA (CEVHDEWMA) chart for monitoring the mean of a Weibull distributed process under type-I censoring. The structure/concept of the HDEWMA chart is equivalent to the TEWMA chart.  

We note here that the generally weighted moving average (GWMA) chart of \cite{Sheu:Lin:2003} is a generalization of the EWMA chart.  Generally, \cite{Mabu:EtAl:2021} provided an extensive review on GWMA control charts. These charts were studied by \cite{knoth2021case} and shown to have no advantages over the much simpler EWMA chart.

Nevertheless, a recursive use of the GWMA statistic was also proposed and used in different applications in the literature. For example,  \cite{Sheu:Hsie:2009} proposed a double version of the GWMA chart; namely the DGWMA chart, for detecting mean shifts. \cite{Chiu:Sheu:2008} and \cite{Chiu:Lu:2015} evaluated the DGWMA chart under Poisson-distributed processes, while \cite{Chen:2020} developed a DGWMA for monitoring COM-Poisson distributed processes. \cite{Huan:Tai:Lu:2014} developed a sum of squares DGWMA (SS-DGWMA) chart. \cite{Alev:Kouk:Lapp:2019} proposed a one-sided DGWMA chart for monitoring TBE data (gamma-distributed). \cite{Lu:2018} proposed a sign-based non-parametric DGWMA chart when the process distribution is unknown, while \cite{Kara:Huma:Niek:2019} developed an exceedance test based DGWMA chart. 

To illustrate the flaws of these methods, we consider the performance of the the DEWMA and TEWMA charts in Section \ref{sec:dewma}.

\subsection{Memory-type Control Charts with Run Rules} \label{subsec:runsrules}

\cite{Abba:Riaz:Does:2011a}
were the first to introduce the use of run rules with a memory-type control chart. They proposed two run rules schemes to be applied on the EWMA chart.
\cite{Abba:EtAl:2015} noted some mistakenly calculated performance measures in \cite{Abba:Riaz:Does:2011a} and provided the correct figures. By this correction, they noted a decrease in the strength of this proposed chart from that reported in \cite{Abba:Riaz:Does:2011a}.
\cite{Khoo:EtAl:2016} extended the study of \cite{Abba:Riaz:Does:2011a} by applying a Markov chain procedure to compute the performance measures. They suggested further run rules schemes than those proposed in \cite{Abba:Riaz:Does:2011a}.
\cite{Mara:Cast:Khoo:2019} derived double integral equations for computing the performance measures of an EWMA chart with run rules. 
\cite{Arsh:EtAl:2017} proposed the simultaneous use of runs rules (nine different schemes) and auxiliary information when monitoring the location parameter using an EWMA chart.

\cite{Riaz:Abba:Does:2011}
proposed two run rules schemes for the CUSUM charts.
\cite{Abba:Riaz:Mill:2012} added some run rules to a scale CUSUM chart designed to monitor the process variability.
\cite{Adeo:Male:2020} considered adding some supplementary run rules to the DEWMA chart design structure for monitoring the process mean.
We consider the use of run rules with the CUSUM and EWMA charts in Section \ref{sec:rr_ewma_cusum}. Again we show that these complications add no performance benefits.

\subsection{Moving Average Methods} \label{subsec:ma}

A moving average (MA) control chart is based on the average of the last specified number of observations   \citep{Robe:1966}. The MA chart is known to be less sensitive than the EWMA and CUSUM charts for small shift sizes \citep{Wong:Gan:Chan:2004,Alev:EtAl:2020}. \cite{zhang2004note} noted that the derivation of the ARL formula of the MA chart provided in the literature -- especially by \cite{wetherill1991statistical} -- is incomplete, and hence the formula is incorrect. \cite{zhang2004note} showed that this incorrect formula could just lead to an upper bound for the ARL value. They derived another formula, but it was not exact. It just provides a sharper upper bound for the ARL value. In their opinion, there is no exact formula that can be found for the ARL metric of the MA charts.

Similar to the DEWMA chart construction, \cite{Khoo:Wong:2008} introduced a double MA (DMA) chart. \cite{Alev:EtAl:2020} noted that the derivation of the DMA chart statistic variance proposed by \cite{Khoo:Wong:2008} is incorrect and provided the correct formula. They re-evaluated the chart as well with the corrected variance. They concluded that the DMA chart is more efficient than the EWMA and CUSUM charts only for the large shift sizes. 

Several evaluations were conducted on the DMA chart. For example,  \cite{areepong2011analytical} and \cite{sukparungsee2014exact} evaluated it under binomial processes, \cite{sukparungsee2013average} under ZIP processes, and \cite{areepong2015explicit} and \cite{areepong2016statistical} under ZIB processes. \cite{phantu2016explicit} designed it for monitoring a ZIB model when the underlying distribution is the ratio of two Poisson means. \cite{phantu2018dma} extended the latter study to develop a DMA chart to monitor Poisson processes modeled by an INAR(1) model, while \cite{raweesawat2021dma} provided some explicit formulas for calculating its ARL when designed to monitor a ZIP-INAR(1) model. \cite{areepong2021double} considered a DMA chart while monitoring zero-truncated Poisson processes. 

\cite{adeoti2019monitoring} proposed the use of a DMA chart to monitor process variability using the sample standard deviation. \cite{Amir:EtAl:2021} studied the use of the chart when there is auxiliary information. Recently, \cite{Alev:Chat:Kouk:2021e,Alev:Chat:Kouk:2021d} proposed triple MA (TMA) and quadruple MA (QMA) charts, respectively. Yet, these compound charts give increasingly more weight to past data values relative to the current data values than the DMA chart, making their use even more inadvisable. We consider the DMA chart performance in Section \ref{sec:madmatmaqma}.

\subsection{The Progressive Mean Approach} \label{subsec:pm}

Progressive mean (PM) charts, originally proposed by
\cite{Abba:EtAl:2012}, are based on the average of all data values obtained during process monitoring. Thus, each data value is given equal weight. \cite{Abba:2015} suggested that a PM chart statistic can be looked at as a special case of the EWMA statistic with an adaptable smoothing parameter (reciprocal of the sample number); and hence can be treated as the Adaptive EWMA statistic proposed by \cite{Capi:Masa:2003}. \cite{Zafa:EtAl:2021} negated the argument of \cite{Abba:2015}, and stated that the PM chart can neither be considered a special case of the EWMA chart nor of the AEWMA chart. This is because of the difference between the variances of the chart statistics, and that the weights are being updated only due to the change of the sample number, not the process status.

This approach showed considerable power to detect changes that occurred when the monitoring starts, but, as reviewed in \cite{Knot:EtAl:2021a}, it grossly under-performed when dealing with later changes. It should be noted that the ``belief'' approach of \cite{Nezh:Niak:2010} is also based on the average of all the process data collected. It thus shares the disadvantages of the PM chart. This approach has been used in various scenarios by by \cite{Asla:Bant:Khan:2019}, \cite{Shaw:Asla:Khan:2020} and \cite{Asla:Khan:Jun:2016b, Asla:Khan:Jun:2017}.

On a recursive basis, \cite{Abba:EtAl:2019a} proposed a double PM (DPM) chart; which weights past data values more than current values.
\cite{Riaz:EtAl:2021c} noted that the variance of the DPM chart statistic derived by \cite{Abba:EtAl:2019a} was incorrect due to some missing terms. \cite{Riaz:EtAl:2021c} derived the correct variance and re-evaluated the chart performance. They concluded that the chart based on the correct variance has much better performance than the older one. 

\cite{Alev:Kouk:2020c} proposed a DPM chart to monitor mean shifts in Poisson processes. \cite{Abba:Nazi:Riaz:2021}
noted that the variance of the chart statistic derived by
\cite{Alev:Kouk:2020c} is incorrect and provided a corrected one. \cite{Abba:EtAl:2021a} introduced a sign-test based on nonparametric DPM control charts.
\cite{Alev:Kouk:2021a} proposed the use of DPM chart to monitor TBE data which are modeled by a gamma distribution.

\cite{Ajad:EtAl:2021} developed two multivariate progressive variance (PV) control charts; one uses the trace and another uses the eigenvalues of the variance-covariance matrix. The charts are evaluated under the assumption of known and unknown in-control process parameters.

Performance of the DPM approach is considered in Section \ref{sec:dpm}.

\subsection{Homogeneously Weighted Moving Average Methods} \label{subsec:hwma}

With the homogeneously weighted moving average (HWMA) approach of  \cite{Abba:2018}, all past data values are weighted equally with the most current data value weighted differently.
\cite{Knot:EtAl:2021a} reviewed this approach and showed that its performance shares the disadvantages of the PM approach.

Double recursive versions (DHWMA charts) were proposed independently by
\cite{Abid:EtAl:2020a} and \cite{Adeo:Kole:2020}.
\cite{Alev:Chat:Kouk:2021a} developed another DHWMA chart to improve upon that proposed by \cite{Abid:EtAl:2020a} and \cite{Adeo:Kole:2020}. Their justification was that the chart developed by \cite{Abid:EtAl:2020a} did not exactly imitate the procedure of \cite{Sham:Sham:1992} in developing the DEWMA chart. Because the chart developed by \cite{Adeo:Kole:2020} has an incorrect variance, \cite{Male:Shon:Adeo:2021} provided a corrected formula of the variance of the control chart statistic. \cite{Anwa:EtAl:2021a} applied the DHWMA chart when there is auxiliary information available. \cite{Riaz:EtAl:2021a} developed a nonparametric DHWMA chart using the sign test. \cite{Alev:Chat:Kouk:2021g} developed a nonparametric DHWMA chart proposed by \cite{Alev:Chat:Kouk:2021a} based on the sign test. 

Generally, we note that the DHWMA
\citep{Abid:EtAl:2020a, Riaz:EtAl:2021a} chart and the triple HWMA chart of \cite{Riaz:EtAl:2021b} are simply reparameterizations of the basic HWMA chart of \cite{Abid:EtAl:2020a} and thus superfluous.

\subsection{Mixed Control Charts} \label{subsec:mixed}

In mixed control charts, the chart statistic of one chart is used as input to another control chart. Many of these types of methods have been introduced since there are many types of charts and  many ways of mixing them.

\cite{abbas2013mixed} was the first to propose mixing the EWMA and CUSUM charts; where they proposed using the EWMA statistic as an input in the CUSUM chart statistics. Following, \cite{zaman2015mixed} proposed the inverse version of \cite{abbas2013mixed} chart by which the CUSUM statistics are used as inputs in the EWMA chart statistic. Generally mixing these two charts has been used extensively. For example, \cite{zaman2016mixed} used this mix in monitoring process dispersion, and \cite{zaman2017performance} used it in monitoring the process location and dispersion simultaneously. \cite{ajadi2016increasing} incorporated headstarts and runs rules into it. A comparison study of these mixed methods was reported by \cite{nazir2016comparative}. \cite{abid2018control} investigated the in-control robustness of this mix of the EWMA and CUSUM under non-normal and contaminated processes. \cite{hussain2020class} proposed a median version for process location monitoring. \cite{Anwa:EtAl:2020,Anwa:EtAl:2021} mixed the two charts with the incorporation of auxiliary information while \cite{mohamadkhani2020developing} considered other sampling scenarios. \cite{aslam2016mixed} mixed these charts for monitoring Weibull-distributed data. \cite{Male:Rapoo:2017} developed a distribution free version, and \cite{osei2017mixed} extended the approach to autocorrelated data. \cite{abbas2018designing} proposed mixing an EWMA statistic with a dual CUSUM chart, \cite{riaz2017mixed} mixed the Tukey EWMA and CUSUM charts, and more recently, \cite{haqnovel2021a} provided another way to combine the two charts. \cite{ajadi2017mixed}, \cite{riaz2019multivariate}, and \cite{Zama:EtAl:2020} extended mixing the control charts to a multivariate level. We study the performance of mixing the EWMA and CUSUM charts in Section 3.

Other mixed control charts were also proposed in the literature. \cite{khan2016ewma}, \cite{ahmad2017double}, \cite{Tabo:Sukp:Aree:2019}, \cite{sukparungsee2020exponentially} and \cite{aslam2021insight} mixed the MA and EWMA charts. In this mix, either the MA statistic is used as input in the EWMA statistic, or inversely the EWMA values are used in the calculation of the MA statistic. \cite{aslam2017double} mixed the DMA and the EWMA charts. Following \cite{Lu:2017}, \cite{ali2018mixed}, \cite{ali2018new} and \cite{Huan:Lu:Chen:2020} mixed the GWMA and CUSUM charts for monitoring the process mean and variance. \cite{mabude2020distribution} developed a nonparametric scheme for the two mixed versions from GWMA and CUSUM charts. \cite{aslam2018hewma} mixed the DEWMA and CUSUM charts to monitor Weibull distributed processes, while \cite{nazir2021efficient} mixed them under normal, heavy-tailed, and skewed process distributions. \cite{taboran2021design} designed a Tukey mixed MA and DEWMA chart. \cite{abid2021mixed,abid2021mixedb} mixed the CUSUM and HWMA charts in both possible orders.

In order to gain the advantage of robustness and efficient detection of small shifts, \cite{abbas2020developing} introduced an EWMA chart under a progressive setup. In this chart, the EWMA and PM charts are integrated such that the PM chart statistic accumulates the EWMA statistics over time instead of the usual sample means. They evaluated the chart under normal and many non-normal distributions. \cite{alevizakos2021developing} noted that the derived chart statistic variance of \cite{abbas2020developing} is incorrect and provided the corrected formula. \cite{ajadi2020progressive} investigated the robustness of this mixed-type chart, and \cite{Ali:EtAl:2021,ali2021designing} proposed a nonparametric version to it. \cite{riaz2021designing} proposed the use of the PM statistic as an input into the EWMA chart statistic.

To illustrate the performance of these approaches, the mixed EWMA-CUSUM, or MEC, chart from \cite{abbas2013mixed} is considered in Section \ref{sec:mec}. The complications of these types of compound charts do not lead to performance advantages over well-designed traditional methods.

\section{Basic Definitions, ARL types and the standard competitor} \label{sec:basics}

For simplicity, we consider an independent series $X_1, X_2, \ldots$ following a normal distribution with mean $\mu$ and standard deviation $\sigma$.
Moreover, the following change point ($\tau$) model
\begin{equation}
  \mu = \begin{cases} \mu_0 = 0 & ,\; t < \tau \\ \mu_1 = \delta & ,\; t \ge \tau \end{cases} \label{eq:tau}
\end{equation}	
is applied. The standard deviation is assumed to be known, $\sigma = \sigma_0 = 1$ (otherwise normalize the $X_t$), and to remain constant.

We denote by $L$ (later it becomes more explicit) the run length stopping time, which is the number of observed $X_i$ values until
an alarm is flagged. Its averages for the two situations, $\tau = 1 $ and $\tau = \infty$, are referred to as
the zero-state Average Run Length (ARL) values, cf. to \cite{Page:1954c, Cros:1986}. Commonly control charts are designed to
exhibit a certain in-control ARL, namely $E_\infty(L) = A$ for some suitably large number $A$.
Afterwards, specific out-of-control ARL values are determined, namely $E_1(L)$ for specified values of $\delta$,
to fill tables or to create ARL profile diagrams.

Focusing to the simple case $\tau = 1$ in \eqref{eq:tau} is often misleading. Therefore, we consider as well the
conditional expected delay (CED)
\begin{align*}
	D_\tau & = E_\tau\big(L-\tau+1\mid L\ge \tau \big)
	\intertext{and, if appropriate, the conditional steady-state ARL}
	\mathcal{D} & = \lim_{\tau\to\infty} D_\tau \,.
\end{align*}
Note that both the sequence of CED values $\{D_\tau\}$ and the limit $\mathcal{D}$ are functions of the shift size $\delta$. For most of the conventional control charts, the sequence converges rapidly to $\mathcal{D}$. Therefore, the conditional steady-state ARL $\mathcal{D}$ is another valuable and representative performance measure. 

For the most part, we use the standard EWMA chart as our benchmark scheme, as proposed by \cite{Robe:1959}.
We use the version with exact control limits, that is, time varying ones.
 \cite{Macg:Harr:1990} acknowledged that these limits introduce some fast initial response properties. So we apply
\begin{align}
	Z_0 & = \mu_0 = 0\quad,\;
	Z_i = (1-\lambda) Z_{i-1} + \lambda X_i \quad,\; i = 1, 2, \ldots \,, \label{eq:ewmaseries} \\
	L_E & = \min \left\{ i\ge 1\!: |Z_i - \mu_0| > c_E \sqrt{ \big( 1-(1-\lambda)^{2i} \big) \frac{\lambda}{2-\lambda} } \,\right\} \,. \label{eq:ewmarule}
\end{align}
In contrast to most of the schemes we review and study in our paper,
numerical routines are fully established to calculate the zero-state ARL, the CED, and the steady-state ARL.
Here, we utilize the \textsf{R} package \texttt{spc} \citep{K:2021}.

\section{Mixed EWMA-CUSUM charts} \label{sec:mec}

After some early experiments applying runs rules to the EWMA chart \citep{Abba:Riaz:Does:2011a}
and to the CUSUM chart \citep{Riaz:Abba:Does:2011}, as discussed in the next section, these authors introduced in \cite{abbas2013mixed}
an amalgam of the EWMA and CUSUM charts, in short the MEC chart. In particular, they consider the usual, see \eqref{eq:ewmaseries}, 
EWMA sequence for $i = 1, 2, \ldots$,
\begin{equation*}
  Q_i = (1-\lambda_q) Q_{i-1} + \lambda_q X_i \,,\;	Q_0 = 0
\end{equation*} 
and utilize $Q_i$ instead of the original $X_i$ as input for a CUSUM chart, namely
\begin{align*}
  M_i^+ & = \max \big\{0, M_{i-1}^+ + Q_i - a_i \big\} \,,\; M_0^+ = 0  \,, \\
  M_i^- & = \max \big\{0, M_{i-1}^- - Q_i - a_i \big\} \,,\; M_0^- = 0  \,.
\end{align*}  
The CUSUM's reference values are set to $a_i = a^* \sigma_{Q, i}$, where
the latter symbol represents the standard deviation of the statistic $Q_i$, cf. to \eqref{eq:ewmarule}:
\begin{equation}
  \sigma_{Q, i} = \sqrt{ \frac{\lambda_q}{2-\lambda_q} \big( 1 - (1-\lambda_q)^{2i}\big)} \,. \label{eq:sigQi}
\end{equation}
In a similar way the alarm rule is adapted, that is,
\begin{equation*}
  L_{MEC} = \min \big\{i\ge 1\!: \max\{M_i^+, M_i^-\} > b^* \sigma_{Q,i} \big\} \,.
\end{equation*}
For given $\lambda_q$, $a^*$ values and some in-control ARL level $A$, the threshold constant $b^*$
is calculated by running a Monte-Carlo study. For $\lambda_q \in \{0.1, 0.25, 0.5, 0.75\}$,
the out-of-control zero-state ARL for selected shifts is determined, again by performing
Monte-Carlo experiments. The authors chose $a^* = 0.5$ and compared the MEC ARL profiles
with the one resulting from a standard CUSUM chart configured with $k = a^*$.
Recall the simple setup of the classical CUSUM:
\begin{align}
  C_i^+ & = \max \big\{0, C_{i-1}^+ + X_i - k \big\} \,,\; C_0^+ = 0 \,, \label{eq:Cplus} \\
  C_i^- & = \max \big\{0, C_{i-1}^- - X_i - k \big\} \,,\; C_0^- = 0 \,, \label{eq:Cminus} \\
  L_C & = \min \big\{i\ge 1\!: \max\{C_i^+, C_i^-\} > h \big\} \,. \label{eq:Carule}
\end{align}
It is not surprising that MEC chart with all the considered $\lambda_q$ values
performs better for small shifts. Here, we propose to compare the MEC design for a
given $\lambda_q$ with a CUSUM chart, where the
$k$ is chosen more appropriately. By using the
asymptotic standard deviation of $Q_i$, the limit of \eqref{eq:sigQi}
\begin{equation*}
  \sigma_{Q, \infty} = \lim_{i\to\infty} \sigma_{Q,i} = \sqrt{ \frac{\lambda_q}{2-\lambda_q} } \,,
\end{equation*}
we set $k = a^* \sigma_{Q, \infty}$. Note that the determination of $h$ is much simpler than that of $b^*$.
After applying this $k = k(\lambda_q, a^*)$ rule, we obtained the 
reference values in Table~\ref{tab:kFORlambda}.
\begin{table}[hbt]
\centering
\caption{Suitable reference values $k$ for the ARL competition between MEC and CUSUM;
alarm thresholds $h$ for the simple CUSUM (in-control $A=170$) are given as well.}
\label{tab:kFORlambda}
\begin{tabular}{cccccccc} \toprule
  $\lambda_q$ &&  0.1 & 0.25 & 0.5 & 0.75 && 1 \\ \midrule
  $k$   &&  0.1147 &  0.1890 & 0.2887 & 0.3873 && 0.5 \\
  $h$   &&  9.8345 &  7.7120 & 5.9798 & 4.8799 && 4.0133 \\ \bottomrule
\end{tabular}
\end{table}
Note that the trivial choice $\lambda_q = 1$ refers to the simple CUSUM chart. The
values $h$ are calculated by using the function \texttt{xcusum.crit()} from
the \textsf{R} package \texttt{spc}.

\begin{figure}[hbt]
\centering
\renewcommand{\tabcolsep}{0mm}
\begin{tabular}{cc}
  \includegraphics[width=.5\textwidth]{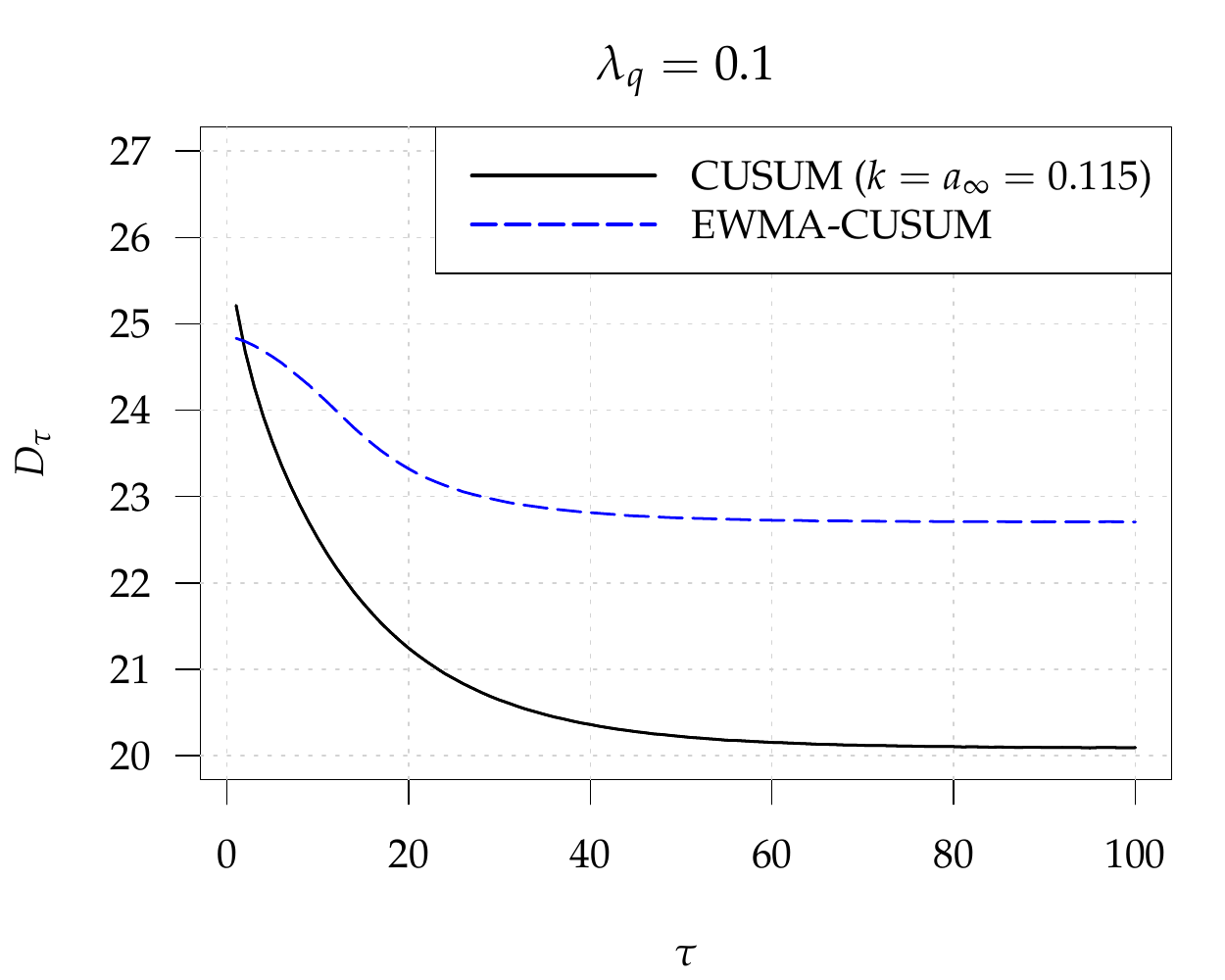} &
  \includegraphics[width=.5\textwidth]{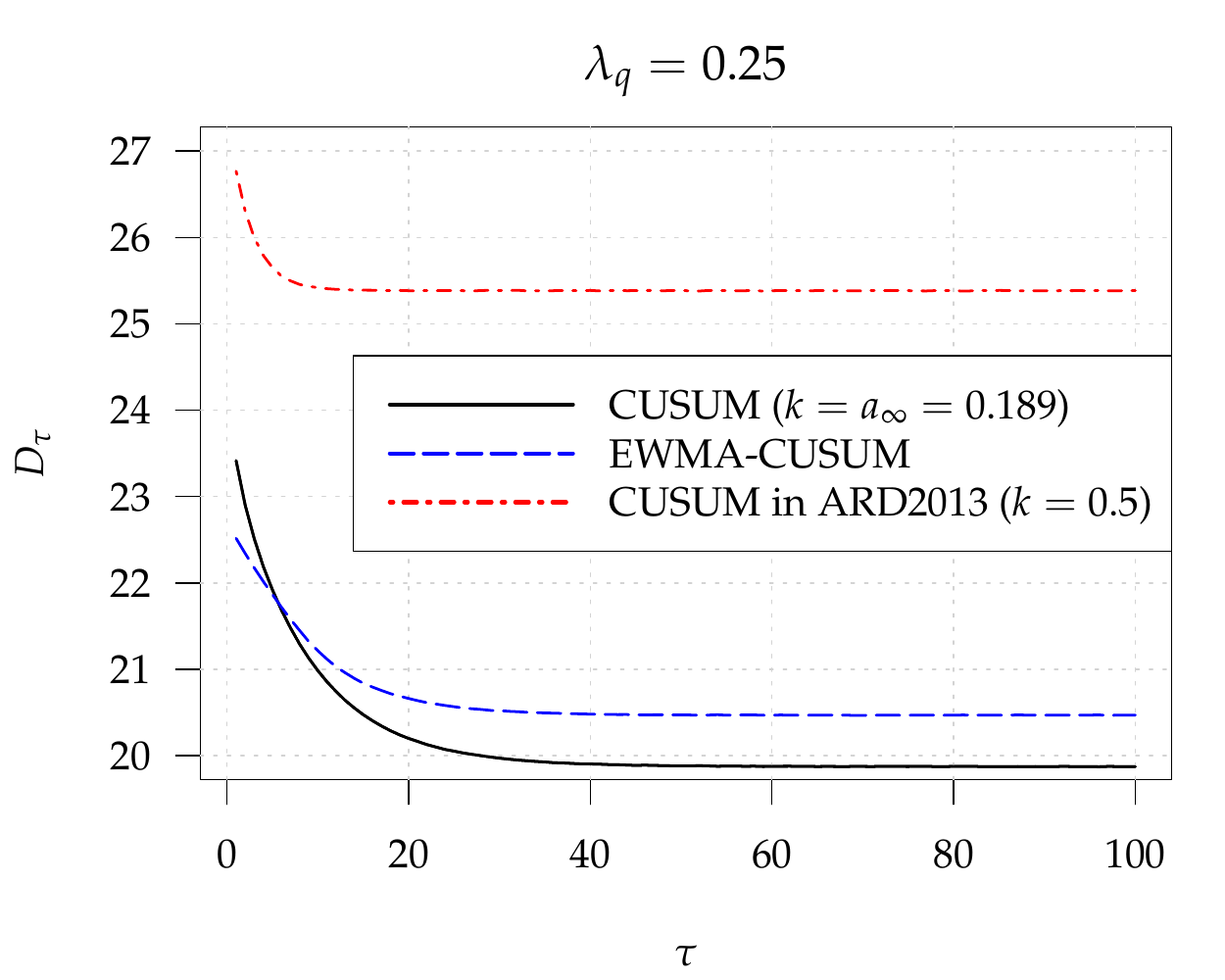}
\end{tabular}
\caption{CED profiles, $D_\tau = E_\tau(L-\tau+1\mid L\ge \tau)$ vs. change-point position $\tau$,  for shift $\delta=0.5$, in-control (zero-state) ARL is $A = 170$;
the red dash-dotted line in the $q=0.25$ diagram refers to the profile $k=0.5$ utilized by
\cite{abbas2013mixed}.}
\label{fig:dtau05MEC}
\end{figure}

For the ARL competition, we look again at the zero-state ARL and the CED with
its limit, the conditional steady-state ARL. In the following figures, we
illustrate the CED profiles for two selected shifts, $\delta \in \{0.5, 1.5\}$.
We start with the case $\delta = 0.5$ in Figure~\ref{fig:dtau05MEC}.
For the considered designs, the MEC chart features a slightly
better zero-state (i.e., $\tau = 1$) ARL values and slightly lower $D_\tau$ values for values of $\tau\le 5$. However, one can conclude that the more complicated MEC scheme
does not exhibit better detection performance than the CUSUM chart. Finally we observe that
the reference value $k=0.5$ (the respective profile is plotted
in the $\lambda_q=0.25$ diagram) chosen in \cite{abbas2013mixed} creates an unfair
disadvantage for the family of CUSUM charts.

\begin{figure}[hbt]
\centering
\renewcommand{\tabcolsep}{0mm}
\begin{tabular}{cc}
  \includegraphics[width=.5\textwidth]{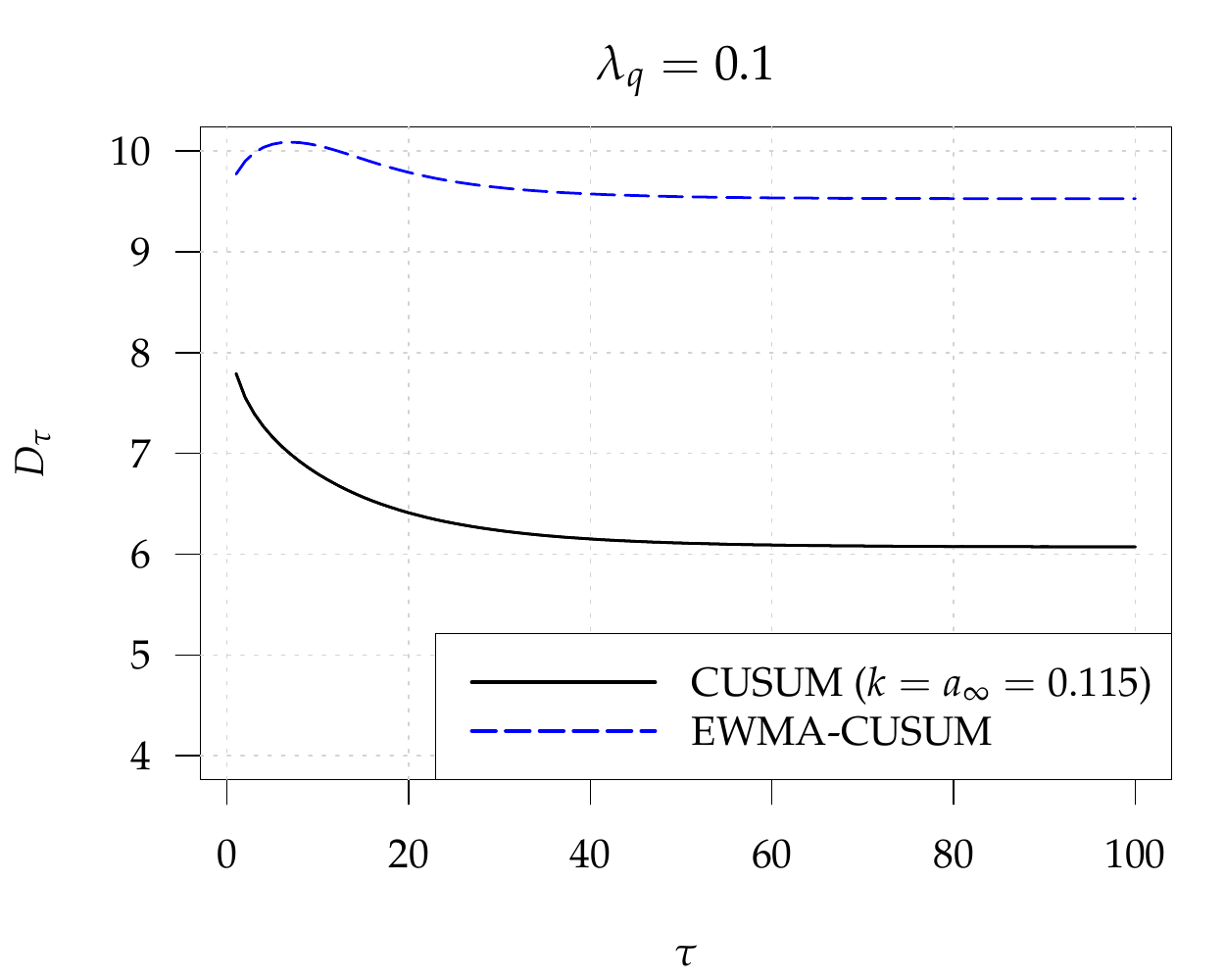} &
  \includegraphics[width=.5\textwidth]{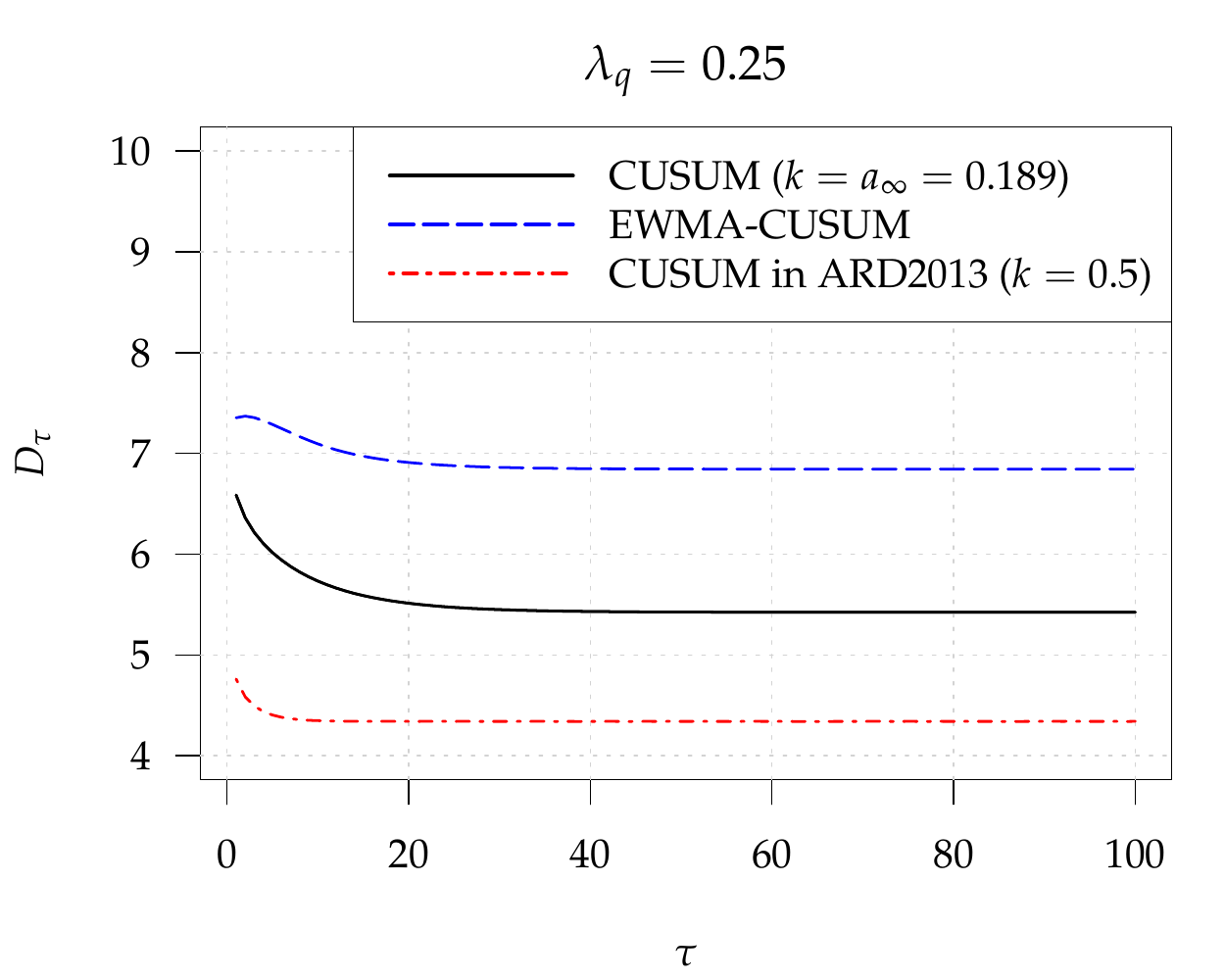}
\end{tabular}
\caption{CED profiles, $D_\tau = E_\tau(L-\tau+1\mid L\ge \tau)$ vs. change-point position $\tau$,  for shift $\delta=1.5$, in-control (zero-state) ARL is $A = 170$;
the red dash-dotted line in the $q=0.25$ diagram refers to the profile $k=0.5$ utilized by
\cite{abbas2013mixed}.}
\label{fig:dtau15MEC}
\end{figure}

Turning to the larger shift, $\delta = 1.5$ in Figure~\ref{fig:dtau15MEC},
we conclude that the MEC chart is uniformly dominated by the respective CUSUM chart.
This is not really surprising, because for larger changes the original CUSUM chart
with $k=0.5$ performs the best.

From all profiles in Figures~\ref{fig:dtau05MEC} and \ref{fig:dtau15MEC} we observe that
the CED converges quite quickly to the conditional steady-state ARL.
Therefore, we are free to utilize the latter as a representative delay measure
for roughly all change point positions.

\begin{figure}[hbt]
\centering
\renewcommand{\tabcolsep}{0mm}
\begin{tabular}{cc}
  \scriptsize $\lambda_q=0.1$, zero-state &
  \scriptsize $\lambda_q=0.25$, zero-state \\[-3ex]
  \includegraphics[width=.5\textwidth]{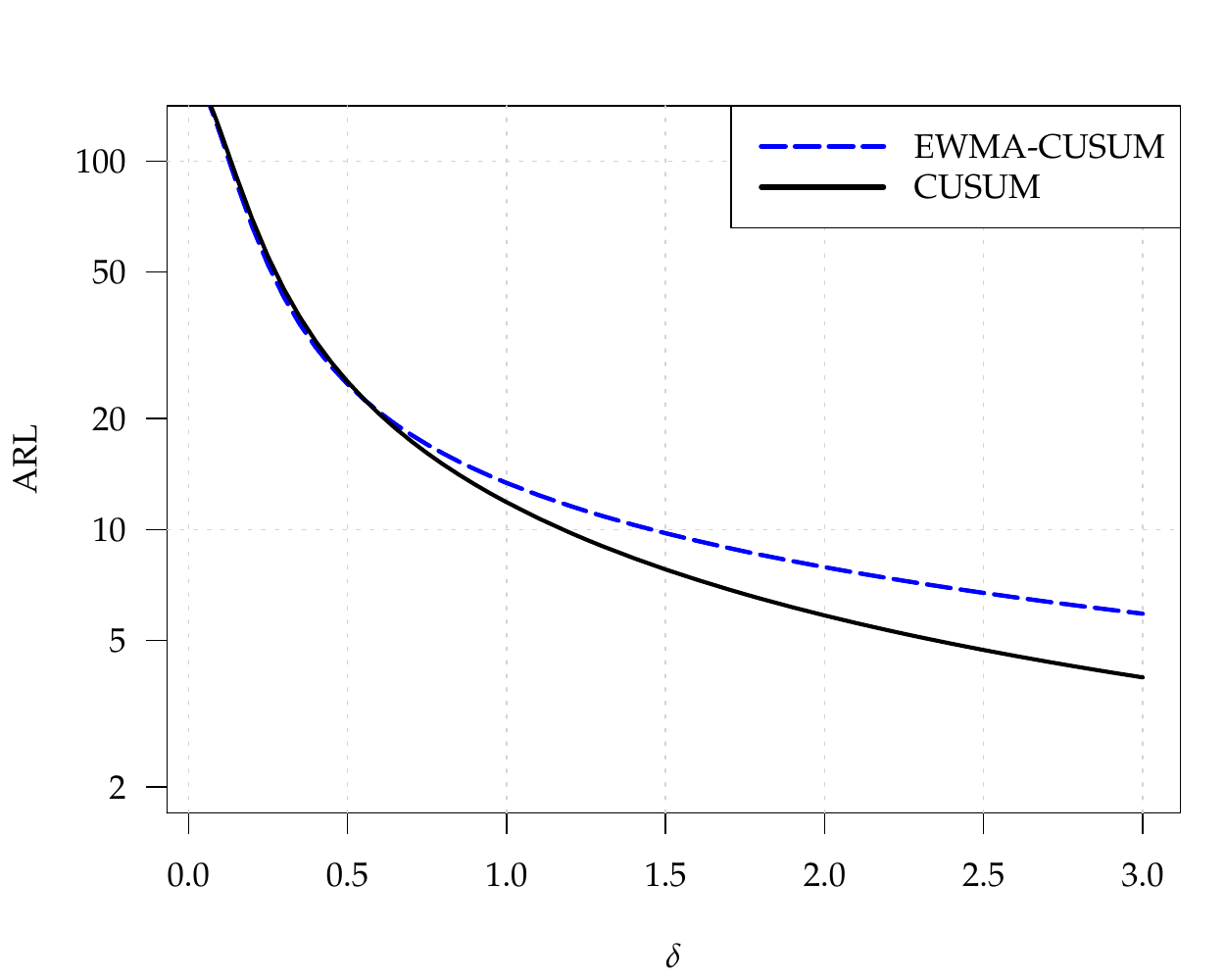} &
  \includegraphics[width=.5\textwidth]{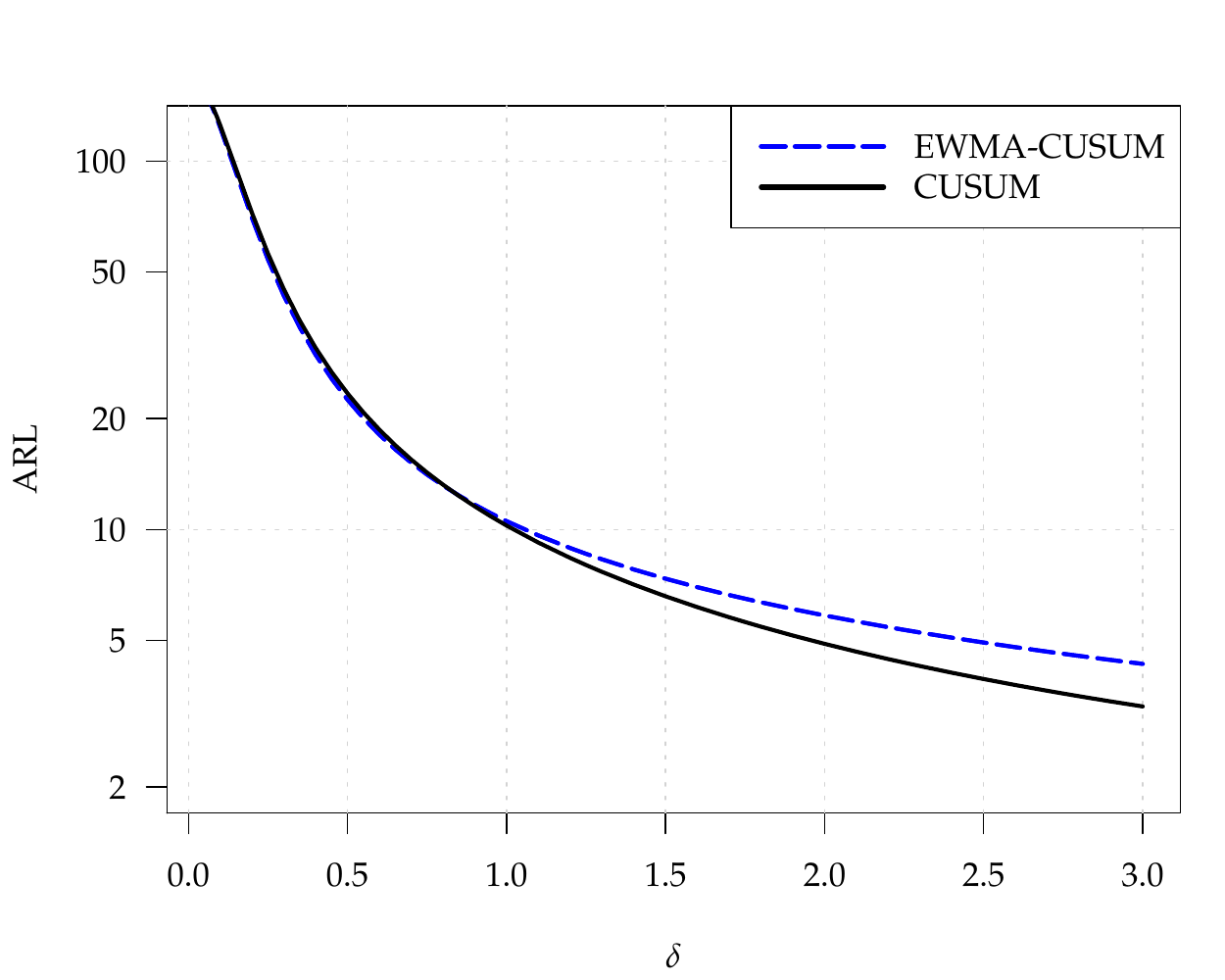} \\[1ex]
  \scriptsize $\lambda_q=0.1$, steady-state &
  \scriptsize $\lambda_q=0.25$, steady-state \\[-3ex]
  \includegraphics[width=.5\textwidth]{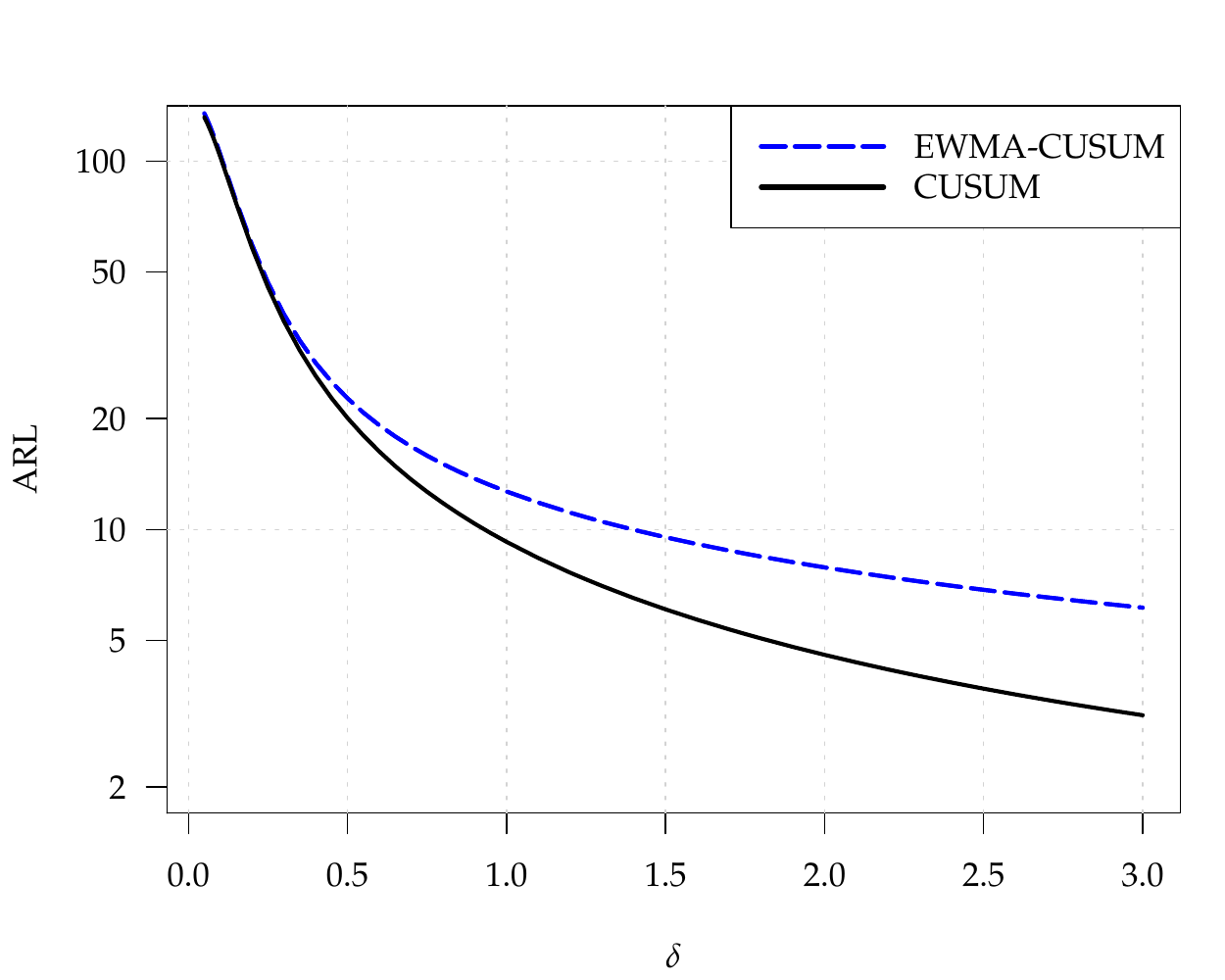} &
  \includegraphics[width=.5\textwidth]{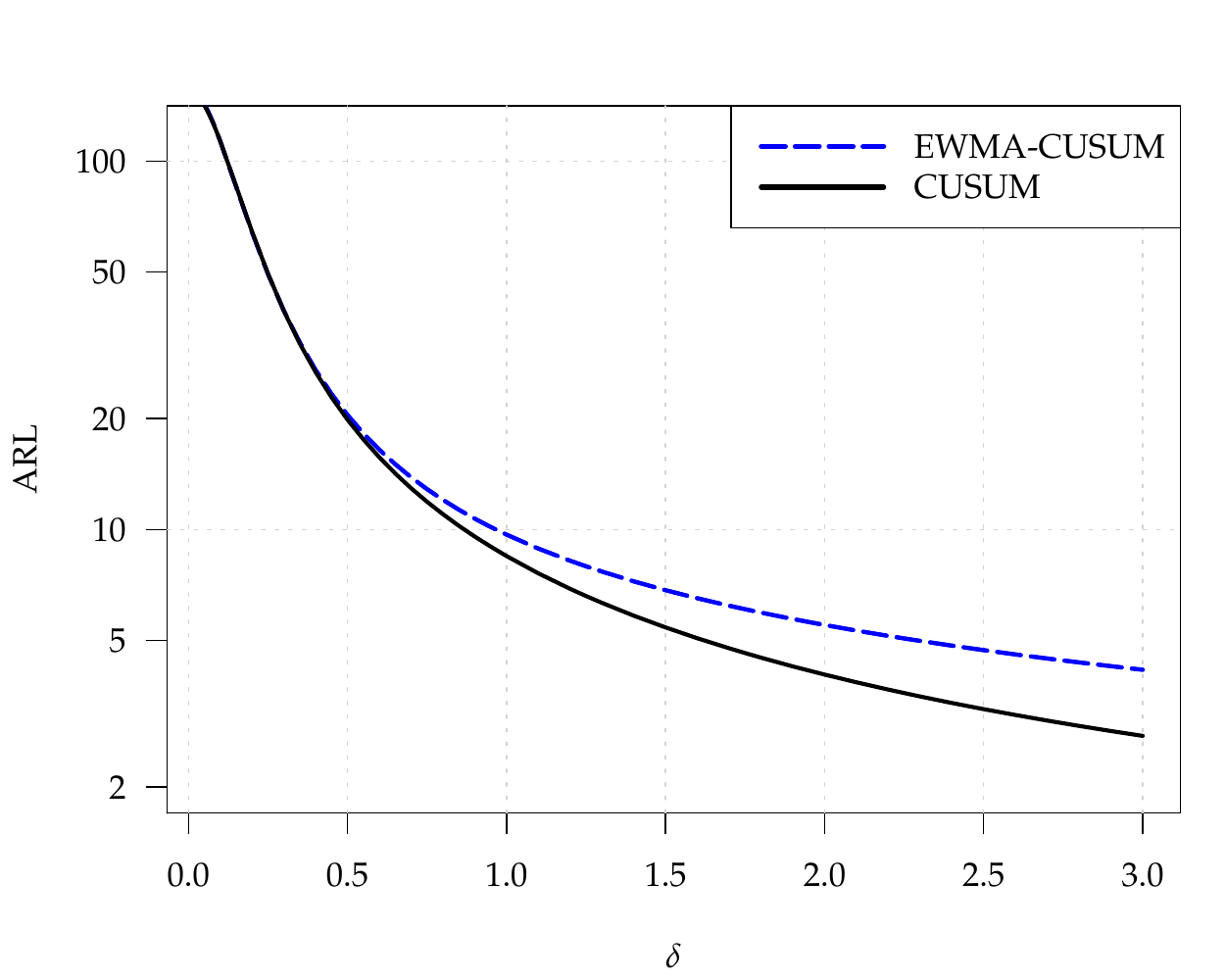}
\end{tabular}
\caption{Zero-state and (conditional) steady-state ARL of MEC and CUSUM
vs. shift size $\delta$, in-control (zero-state) ARL is $A = 170$.}
\label{fig:arlMEC}
\end{figure}

In Figure~\ref{fig:arlMEC} we present a final ARL analysis of our two MEC designs
($\lambda_q = 0.1$ and $= 0.25$)
together with their CUSUM chart opponent.
Essentially, the simpler CUSUM chart exhibits better or similar ARL results compared to the MEC chart. For very small $\delta$ values, the MEC chart exhibits slightly lower out-of-control
zero-state ARL results, whereas for medium size and large shifts, the CUSUM chart is
substantially better. In case of the steady-state ARL, the CUSUM chart is for all
shifts as least as good as the MEC chart while being much better again for medium size shifts
and larger ones.

\begin{figure}[hbt]
\centering
\renewcommand{\tabcolsep}{0mm}
\begin{tabular}{cc}
  \scriptsize $\lambda_q=0.1$ &
  \scriptsize $\lambda_q=0.25$ \\[-1ex]
  \includegraphics[width=.51\textwidth]{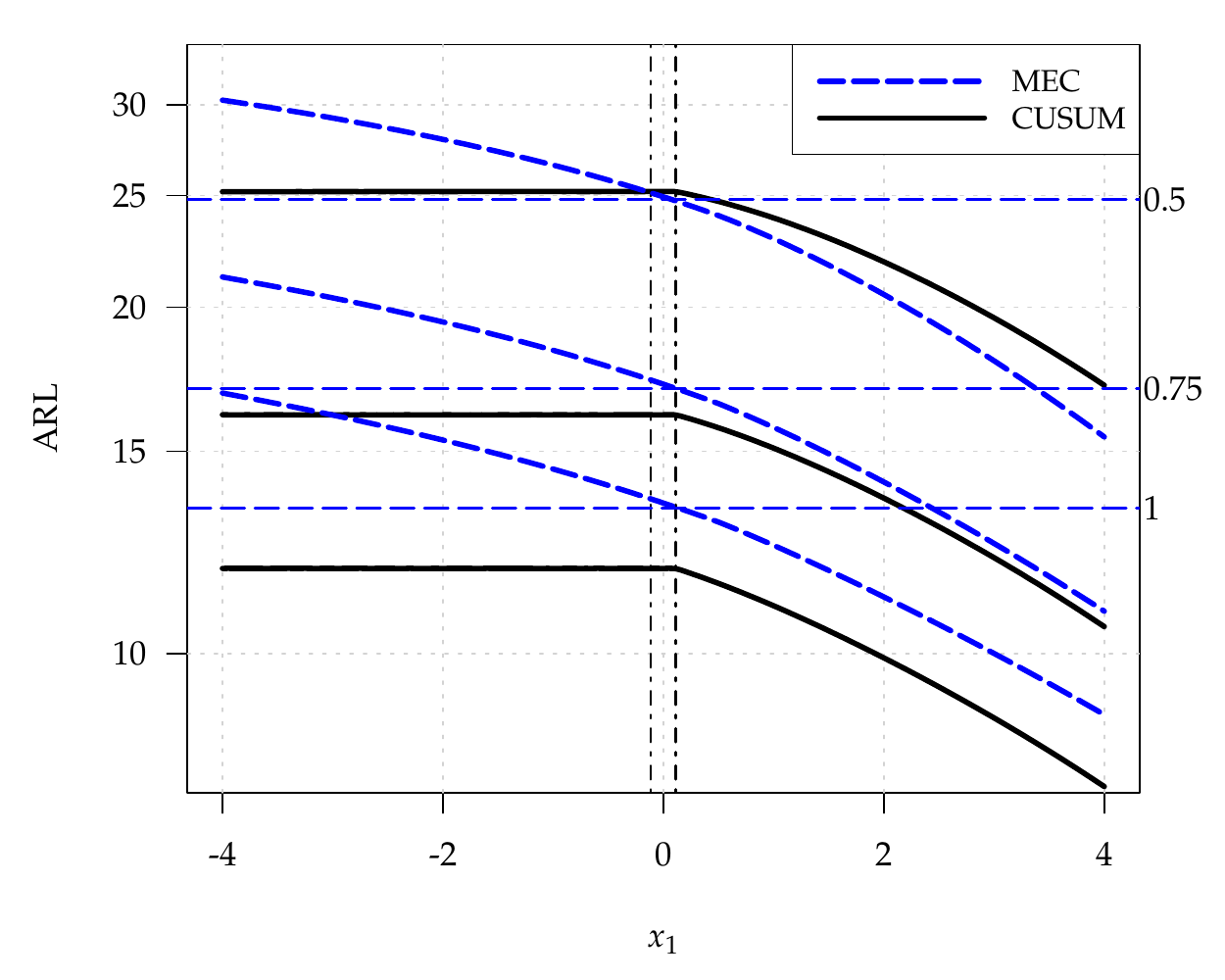} &
  \includegraphics[width=.51\textwidth]{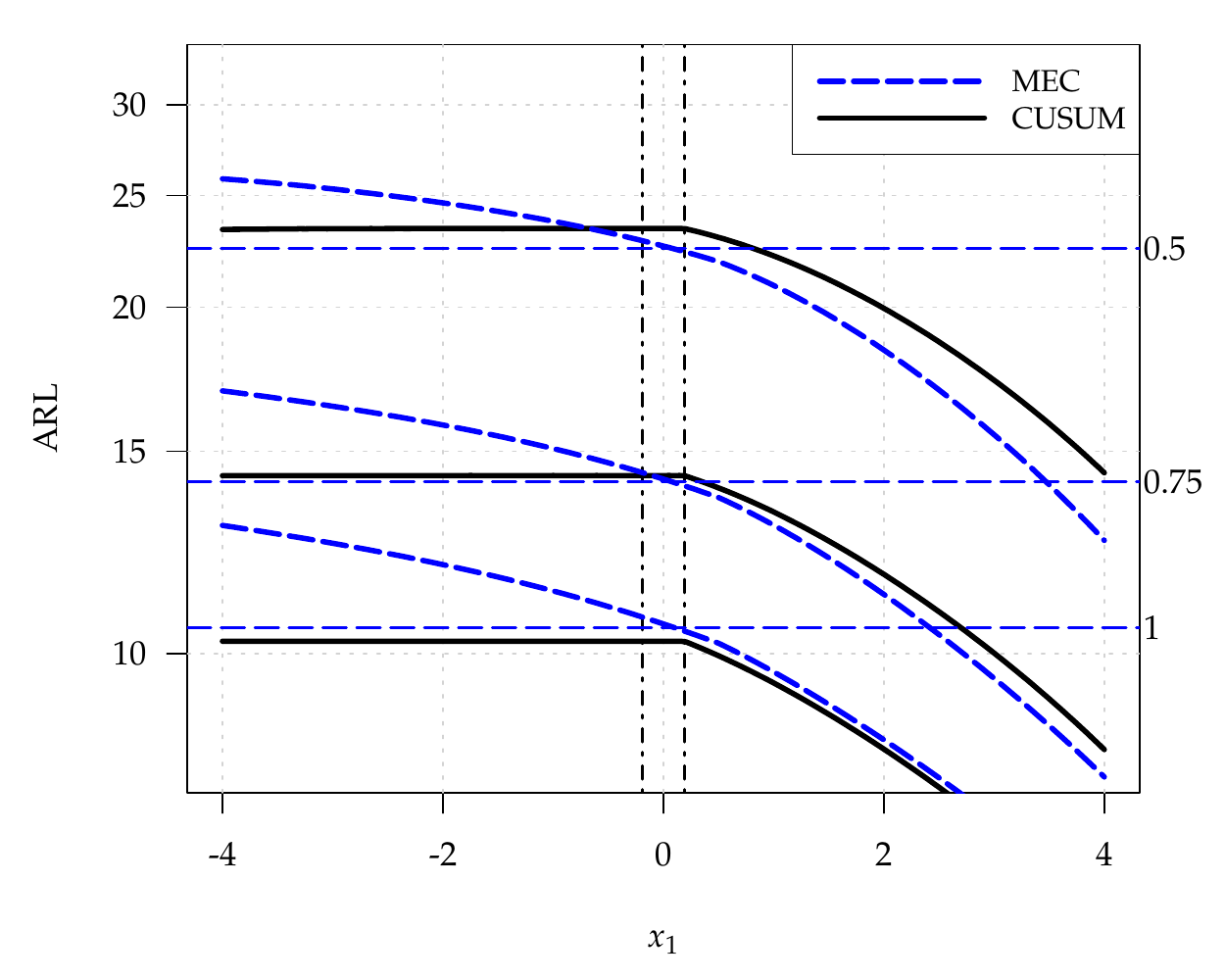} \\
\end{tabular}
\caption{Effective out-of-control ARL, conditioned on first observation
$X_1 = x_1$ for MEC and CUSUM; three shifts $\delta \in \{0.5, 0.75, 1\}$;
MEC zero-state ARL added as horizontal lines;
vertical lines are added at $(-k,k)$.}
\label{fig:MECworst}
\end{figure}

Finally we want to mention that in the case of the CUSUM chart, the zero-state ARL is equal to
the worst case ARL, where \cite{Mous:1986} proved optimality (at $\delta = 2k$) for the CUSUM chart.
Contrarily, for the MEC chart the worst case ARL is larger than the zero-state ARL. Refer to Figure~\ref{fig:MECworst}.
The worst-case MEC ARL is more difficult to determine, because the worst case is characterized by $M_{\tau-1}^+ = 0$
and $Q_{\tau-1} \ll 0 = \mu_0$, whereas for the zero-state ARL we deploy simply $Q_0 = 0$ (change-point $\tau = 1$). We demonstrate this by considering a simplified
situation. We condition on the first observation $X_1 = x_1$ and calculate
the ARL for $\tau = 2$. Next, we plot these values as a function of
$-4 \le x_1\le 4$ (as long as $|x_1| \le h + k$ and $\le b^* + a^*$, it does not
trigger an alarm, respectively; in all cases the right-hand sides are larger than 5).
We consider three different shifts, $\delta \in \{0.5, 0.75, 1\}$.
First, we recognize the constant parts of the CUSUM profiles for $x_1 \le k$.
Second, the MEC chart branches on the left-hand side reside clearly above the
reported zero-state ARL (and as well above $D_2$). For the CUSUM chart, however,
the constant part is at the zero-state ARL level.
The values in Figure~\ref{fig:MECworst} were estimated using Monte Carlo
simulations. Differently from $D_2$ (plotted in Figures~\ref{fig:dtau05MEC}
and \ref{fig:dtau15MEC}), the first observation $X_1$
is not drawn from $\mathcal{N}(0,1)$ (the in-control model), but set
to be just $X_1 = x_1$. In this way, we simulated
\begin{equation*}
  \ell(x_1) :=
  E_2(L-1\mid L\ge 2, X_1=x_1) =
  E_2(L-1\mid X_1=x_1) \quad \text{for $-4\le x_1\le 4$} \,.
\end{equation*}
And of course $\sup_{-4\le x_1\le 4} \ell(x_1)= \mathcal{L}$ for the CUSUM chart.
In Figure~\ref{fig:MECworst} we observe that the respective $\sup$ for
the MEC chart might be substantially larger than its zero-state ARL $\mathcal{L}$.

Summing up, we find that for any MEC chart design, we can find a CUSUM control chart that exhibits nearly the same zero-state ARL, lower steady-state
ARL and lower worst case ARL. Thus, the larger efforts involved in setting up and using a MEC chart do not pay off. A simpler CUSUM chart does the job better
with less effort.

\section{Runs rule CUSUM and EWMA charts} \label{sec:rr_ewma_cusum}

\cite{Riaz:Abba:Does:2011} and \cite{Abba:Riaz:Does:2011a} proposed to impose on the CUSUM and EWMA charts certain runs rules.
The latter are classical improvements to Shewhart type charts to introduce some memory into these simple
and very popular devices. These rules were considered in \cite{Dudd:Jenn:1942} and \cite{Weil:1953},
but more familiar references are likely \cite{WECO:1956} and \cite{Nels:1984}.
Runs rules experienced some renaissance later on, see, for example, \cite{Klei:2000a}, \cite{Khoo:2003b}
and \cite{Kout:Bers:Mara:2007}. Regarding their ARL analysis we refer to \cite{Cham:Wood:1987} and \cite{Cham:1992}.

In \cite{Riaz:Abba:Does:2011}, classical 2-of-2 and 2-of-3 rules augment a standard CUSUM control chart as
supplementary rules. Differently, \cite{Abba:Riaz:Does:2011a} established the classical 2-of-2 and a modified 2-of-3
rule as sole alarm rule for an EWMA control chart. While the former needs alarm and warning limits (denoted with $AL$ and $WL$),
the latter utilizes only warning thresholds.

We start with the two RR-CUSUM schemes in \cite{Riaz:Abba:Does:2011}.
In both cases, an alarm is triggered if the standard CUSUM rule triggers a signal, see \eqref{eq:Carule}, or
two of two (three) consecutive points of $C_i^+$ or $C_i^-$ in \eqref{eq:Cplus} and \eqref{eq:Cminus}, respectively,
reside between the warning limit $W\!L$ and the alarm limit $AL = h$. In both publications --- cf. to the
corresponding PhD thesis \cite{Abba:2012b} --- the zero-state ARL was estimated using a Monte Carlo
study. By utilizing the Excel Add-In \texttt{MCSim} \citep{Bois:Masz:1997} and 5\,000 replications, all the
ARL estimates were determined. Unfortunately, neither in the journal publications nor in the PhD thesis was the Monte Carlo design sufficiently described in detail. As examples, we picked some results from Table II/III in \cite{Riaz:Abba:Does:2011}
and compared them to results of a more extensive Monte Carlo study with $10^8$ replications in Table~\ref{tab:ARD2011cusumTab23}.
\begin{table}[hbt]
\centering
\caption{ARL results from Table II/III in \cite{Riaz:Abba:Does:2011} augmented
with new results obtained with $10^8$ replications (second line).}\label{tab:ARD2011cusumTab23}
\renewcommand{\tabcolsep}{1.25mm}
\begin{tabular}{ccccccccccc} \toprule
	&&& \multicolumn{6}{c}{$\delta$} && \\[0.3ex]
	$WL$ & $AL$ && 0.25 & 0.5 & 0.75 & 1 & 1.5 & 2 && $AL^*$ \\ \midrule
	\multicolumn{11}{c}{2-of-2} \\ \midrule
	3.42 & 4.8 && 71.87 & 25.56 & 13.54 & 8.66 & 5.08 & 3.68 && $\infty$ \\
	&&& 73.60 & 26.68 & 13.51 & 8.67 & 5.08 & 3.69 &&	\\ \midrule
	3.44 & 4.6 && 72.26 & 25.65 & 13.50 & 8.57 & 5.01 & 3.61 && 4.65 \\
	&&& 73.91 & 26.70 & 13.48 & 8.62 & 5.01 & 3.61 && \\ \midrule
	3.48 & 4.4 && 71.94 & 25.59 & 13.50 & 8.52 & 4.94 & 3.52 && 4.38 \\
	&&& 74.31 & 26.73 & 13.44 & 8.56 & 4.94 & 3.53 && \\
	\midrule
	3.53 & 4.2 && 71.40 & 25.30 & 13.33 & 8.40 & 4.83 & 3.42 && 4.23 \\
	&&& 73.55 & 26.51 & 13.30 & 8.44 & 4.84 & 3.44 && \\ \midrule
	\multicolumn{11}{c}{2-of-3} \\ \midrule
	3.5 & 4.44 && 71.49 & 25.38 & 13.40 & 8.46 & 4.94 & 3.54 && 4.52 \\
	&&& 73.38 & 26.52 & 13.37 & 8.54 & 4.95 & 3.55 && \\ \midrule
	3.6 & 4.19 && 72.94 & 25.37 & 13.35 & 8.38 & 4.83 & 3.42 && 4.18 \\
	&&& 74.13 & 26.61 & 13.33 & 8.46 & 4.84 & 3.44 && \\ \midrule
	3.7 & 4.08 && 73.11 & 25.37 & 13.31 & 8.34 & 4.78 & 3.38 && 4.08 \\
	&&& 74.23 & 26.62 & 13.30 & 8.41 & 4.79 & 3.38 && \\ \midrule
	3.8 & 4.03 && 73.59 & 25.40 & 13.28 & 8.32 & 4.75 & 3.35 && 4.03 \\
	&&& 74.29 & 26.64 & 13.29 & 8.39 & 4.76 & 3.36 && \\
	\bottomrule
\end{tabular}
\end{table}
Moreover, we provided slightly changed alarm limits, $AL^*$, that ensure the nominal in-control zero-state ARL of 168.
Note that $AL^* = \infty$ corresponds to applying only the runs rule. Note that
for all shifts $\delta \ge 0.75$ we clearly confirm the results of \cite{Riaz:Abba:Does:2011}, whereas
we observe a gap between our values and theirs for the smaller shifts, $\delta = 0.25$ and $= 0.5$. Using the corrected alarm
limits, $AL^*$, and the larger ``power'' of a $10^8$ Monte Carlo study, we repeated
the competition between the two RR-CUSUM control charts and the standard CUSUM chart.
\begin{table}[hbt]
\centering
\caption{Repeating the \cite{Riaz:Abba:Does:2011} competition between the 2-of-2,
2-of-3 and the standard CUSUM.}\label{tab:ARD2011cusumTab23NEW}
\renewcommand{\tabcolsep}{1.25mm}
\begin{tabular}{ccccccccc} \toprule
	&&& \multicolumn{6}{c}{$\delta$} \\[0.3ex]
	$WL$/$k$ & $AL$/$h$ && 0.25 & 0.5 & 0.75 & 1 & 1.5 & 2 \\ \midrule
	\multicolumn{9}{c}{2-of-2} \\ \midrule
	3.42 & $\infty$ && 74.12 & 26.88 & 13.68 & 8.84 & 5.31 & 4.00 \\
	3.44 & 4.65     && 74.12 & 26.75 & 13.51 & 8.64 & 5.04 & 3.63 \\
	3.48 & 4.38     && 74.13 & 26.69 & 13.42 & 8.54 & 4.93 & 3.52 \\
	3.53 & 4.23     && \B 74.05 & \B 26.63 & \B 13.35 & \B 8.47 & \B 4.86 & \B 3.45 \\ \midrule
	\multicolumn{9}{c}{2-of-3} \\ \midrule
	3.5 & 4.52 && \B 73.87 & 26.64 & 13.43 & 8.58 & 4.99 & 3.58 \\
	3.6 & 4.18 && 73.96 & \B 26.58 & 13.32 & 8.45 & 4.83 & 3.43 \\
	3.7 & 4.08 && 74.23 & 26.62 & \B 13.30 & 8.41 & 4.79 & 3.38 \\
	3.8 & 4.03 && 74.29 & 26.64 & 13.29 & \B 8.39 & \B 4.76 & \B 3.36 \\ \midrule
	\multicolumn{9}{c}{standard CUSUM} \\ \midrule
	0.50  & 4.002 && 74.31 & 26.65 & 13.29 & 8.39 & 4.75 & 3.34 \\
	0.4933 & 4.045 && \B 73.86 & \B 26.48 & \B 13.25 & \B 8.39 & \B 4.76 & \B 3.36 \\
	0.49  & 4.067 && 73.64 & 26.40 & 13.23 & 8.39 & 4.77 & 3.37 \\
	0.48  & 4.134 && 72.96 & 26.16 & 13.18 & 8.39 & 4.80 & 3.39 \\
	\bottomrule
\end{tabular}
\end{table}
From Table~\ref{tab:ARD2011cusumTab23NEW} we conclude that the $k = 0.4933$ CUSUM chart beats all the RR-CUSUM designs.
But the main result is that there is no convincing reason to abstain from using the initial $k = 0.5$ CUSUM chart, whose
out-of-control ARL results are sufficiently small compared
to all considered RR-CUSUM schemes.

Turning to \cite{Abba:Riaz:Does:2011a} RR-EWMA schemes, we observe several particularities.
As already mentioned, the runs rule is deployed standalone. Second, the 2-of-3 rule is more involved.
The version for the lower limit is given by \textit{``At least two out of three consecutive points fall below an LSL and the point above the LSL (if any) falls between the CL and the LSL''} \citep{Abba:Riaz:Does:2011a}.
Third in \cite{Abba:Riaz:Does:2011a} and as well in \cite{Abba:2012b}, wrong ARL results for the modified
2-of-3 rule were reported. Later, in \cite{Abba:EtAl:2015}, these values were corrected. Presumably, \cite{Khoo:EtAl:2016}
evoked these corrections by writing
\textit{``While highlighting some erroneous average run length (ARL) and standard deviation of the run length (SDRL) computations in Abbas et al. (2011)''}.
At least, \cite{Abba:EtAl:2015}
explained what happened some years before: \textit{``The reason is being the omitted statement in a simulation code (mistakenly)
dealing with the lower sided limit''}. This means that ARL values of a one-sided scheme were reported inadvertently. This led as well to
an incorrect control (alias warning) limit. Fourth, the authors compared the two RR-EWMA schemes with many other control charts,
but not to the standard EWMA chart, cf. to \eqref{eq:ewmarule}. Here, see Table~\ref{tab:ARD2011ewmaTab2}, we start with a comparison of the 2-of-2 EWMA chart with the standard EWMA chart.
\begin{table}[hbt]
\centering
\caption{Table II of \citep{Abba:Riaz:Does:2011a} and standard EWMA ARL results.} \label{tab:ARD2011ewmaTab2}
\renewcommand{\tabcolsep}{1.25mm}
\begin{tabular}{*{13}{r}} \toprule
  && \multicolumn{11}{c}{EWMA w/ and w/o runs rule} \\[0.3ex]
  $\delta$ && \multicolumn{2}{c}{$\lambda=0.1$} && \multicolumn{2}{c}{$\lambda=0.25$} &&
  \multicolumn{2}{c}{$\lambda=0.5$} && \multicolumn{2}{c}{$\lambda=0.75$} \\ \midrule
  0.00  && 169.87 & \B 169.99 && 169.48 & 169.99 && 169.68 & 169.98 && 169.71 & 170.02 \\
  0.25  &&  54.58 & \B  53.98 &&  73.48 &  74.54 &&  94.82 &  99.53 && 110.78 & 118.63 \\
  0.50  &&  19.80 & \B  18.92 &&  26.63 &  26.46 &&  37.99 &  41.20 &&  49.06 &  58.77 \\
  0.75  &&  10.59 & \B  9.79 &&  12.95 &  12.63 &&  17.41 &  19.22 &&  23.29 &  29.29 \\
  1.00  &&   6.94 & \B   6.16 &&   7.95 &   7.48 &&   9.88 &  10.51 &&  13.07 &  15.90 \\
  1.50  &&   4.11 & \B   3.28 &&   4.35 &   3.76 &&   4.73 &   4.51 &&   5.58 &   6.15 \\
  2.00  &&   2.98 & \B  2.16 &&   3.10 &   2.41 &&   3.12 &   2.69 &&   3.34 &   3.21 \\ \midrule
  $L_S$/$c_E$ &&  2.145 & 2.2145 &&  2.184 & 2.6282 &&  2.034 & 2.7241 &&  1.830 & 2.7493 \\
  \bottomrule
\end{tabular}
\end{table}
We match the competitors by simply using the same value for $\lambda$ as done in \cite{Abba:Riaz:Does:2011a}.
For the standard EWMA we used 170 as nominal
in-control ARL. First, we see from Table~\ref{tab:ARD2011ewmaTab2} that the uniformly best out-of-control ARL values are obtained for the standard EWMA chart with $\lambda = 0.1$.
We observe for both EWMA schemes, that the out-of-control ARL decreases with decreasing $\lambda$.
Finally, setting $\lambda$ of the standard EWMA to $= 0.23$, $= 0.44$ and $= 0.61$, we obtain designs whose out-of-control
ARL values are uniformly smaller than the ones of the 2-of-2 EWMA with $\lambda = 0.25$, $= 0.5$ and $= 0.75$, respectively.

Finally, we consider the modified 2-of-3 EWMA chart and focus on one example, namely $\lambda = 0.1$ and target
in-control ARL 168. In Table~\ref{tab:ARD2011ewmaTab3}, we provide the incorrect numbers from \cite{Abba:Riaz:Does:2011a} and
the corrected ones in \cite{Abba:EtAl:2015}, new ones from another Monte Carlo study with $10^8$ replications for the common 2-of-3 EWMA chart and
results for the standard EWMA chart calculated with the \textsf{R} package \texttt{spc}.
\begin{table}[hbt]
\centering
\caption{Zero-state ARL values for modified 2-of-3 EWMA from Table III of \cite{Abba:Riaz:Does:2011a}, \cite{Abba:EtAl:2015} 
and a further Monte Carlo study for the common 2-of-3 EWMA,
and for standard EWMA (\textsf{R} package \texttt{spc}); $\lambda = 0.1$.} \label{tab:ARD2011ewmaTab3}
\renewcommand{\tabcolsep}{1.25mm}
\begin{tabular}{*{9}{r}} \toprule
  && \multicolumn{2}{c}{modified 2-of-3} && 2-of-3 && standard \\[0.3ex]
  $\delta$ && (\citeyear{Abba:Riaz:Does:2011a}) & (\citeyear{Abba:EtAl:2015}) && $10^8$ && \texttt{spc} \\ \midrule
  0.00  && 167.32 & 167.09 && \textit{166.36} && 168.00 \\
  0.25  &&  34.37 &  53.71 &&  \textit{53.70} &&  53.59 \\
  0.50  &&  14.04 &  19.37 &&  \textit{19.40} &&  18.83 \\
  0.75  &&   8.13 &  10.40 &&  \textit{10.42} &&   9.75 \\
  1.00  &&   5.71 &   6.83 &&   \textit{6.84} &&   6.14 \\
  1.50  &&   3.78 &   4.01 &&   \textit{4.02} &&   3.27 \\
  2.00  &&   3.20 &   2.96 &&   \textit{2.96} &&   2.15 \\ \midrule
  $L_S$/$c_E$ &&  1.807 &  2.158 && 2.158 && 2.4098 \\
  \bottomrule
\end{tabular}
\end{table}
At the bottom line we added as well the control limit factors used. We observe again that the standard EWMA chart exhibits the lowest zero-state ARL numbers. Regarding the comparison for other values of $\lambda$, we observe a similar behavior as for the 2-of-2-EWMA discussed above. Comparing the results of \cite{Abba:EtAl:2015} and the ones for the
common 2-of-3 EWMA, we acknowledge their similarity.
It indicates that the modified 2-of-3 rule could be reduced to
the standard one, where the \textit{``point above the LSL (if any)
falls between CL and the LSL''} condition is dropped.

\cite{Khoo:EtAl:2016} provided a Markov chain approximation technique for the two RR-EWMA charts. However, there are two differences. First, they considered constant signal limits relying on the asymptotic EWMA variance $\lambda/(2-\lambda)$. Second, they utilized the standard (they called it non-side-sensitive) 2-of-3 runs rule, where it is not important anymore, whether the ``unobtrusive'' EWMA value is on the right side of the center line.
We dealt with this case in Table~\ref{tab:ARD2011ewmaTab3}
already.
\cite{Khoo:EtAl:2016}
investigated further types of RR-EWMA charts, where
for the most complicated one (four runs rules) it was declared \textit{``it is found that generally the four runs rules EWMA
schemes outperform all charts under comparison, in terms of ARL''}. However, they applied again the oversimplified approach of equalizing the $\lambda$ values. If we compare the four runs rule EWMA charts with $\lambda = 0.1, 0.2, 0.3, \ldots, 0.9$ to the standard EWMA chart with instead $\lambda = 0.09, 0.18, 0.26, \ldots, 0.62$,
respectively,
then we obtain charts which exhibit lower out-of-control ARL values for all considered
\citep[in][]{Khoo:EtAl:2016} shifts
$\delta = 0.1, 0.2, \ldots 7$.
In summary, the ad hoc use of runs rules EWMA charts is not
worth the required effort.

\section{MA, DMA, TMA and QMA Charts} \label{sec:madmatmaqma}

The moving average (MA) is a well-known filtering technique for
time series data. Thus, it is not surprising that it was proposed
as well for doing process monitoring, but only relatively few papers have dealt with it.
\cite{Robe:1966} was presumably the first, whereas \cite{Lai:1974}
provided some results for moving averages in general.
Then \cite{Wong:Gan:Chan:2004} provided some guidelines
to apply MA control charts. Essentially, the latter authors
found that the detection performance of MA charts to be quite similar to that of EWMA charts. They claimed as well that the MA
principle is  more popular than EWMA in several
fields such as finance. However, later on
\cite{Khoo:Wong:2008} introduced the so-called double MA (DMA) chart.
It is based on the following equations with $w$ representing the window size:
\begin{equation*}
  M_i = \begin{cases}
    \frac1i \sum_{j=1}^i X_j & ,\; i\le w \\
    \frac1w \sum_{j=i-w+1}^i X_j & ,\; i\ge w \\
  \end{cases} \qquad , \quad
  D_i = \begin{cases}
    \frac1i \sum_{j=1}^i M_j & ,\; i\le w \\
    \frac1w \sum_{j=i-w+1}^i M_j & ,\; i\ge w \\
  \end{cases} \quad .
\end{equation*}
Beginning with $i\ge w$,
the ordinary MA statistics $M_i$ sum the most recent $w$ observations with equal weight $1/w$. In contrast, the double MA statistic involves more data points.
Namely starting with $i\ge 2w -1$ it sums the most
recent $2w-1$ observations with triangular weights
$1/w^2, 2/w^2, \ldots, (w-1)/w^2, w/w^2, (w-1)/w^2, \ldots, 1/w^2$.
Obviously, the data point in the middle of the window is given the same
weight as the ones in the MA sum, whereas the others get smaller weights.
Straightforwardly, \cite{Khoo:Wong:2008} proposed the
control limits $\mu_0 \pm L \sqrt{Var(D_i)}$.
However, they made a mistake while calculating the variance
term. \cite{Alev:EtAl:2020} spotted it and provided correct
formulas. Here, we want to consider only the case $i \ge 2w-1$,
where \cite{Khoo:Wong:2008} gave an overly simple result.We have
\begin{small}
\begin{align*}
  Var(D_i) & = \begin{cases}
    \displaystyle\frac{1}{w^2}
    & \!\!\text{, \cite{Khoo:Wong:2008}, } \\[3.2ex]
    \displaystyle\frac{1}{w^2} \left[ 1 + \frac{2}{w^2} \sum_{w\le j_1 < j_2 \le 2w-1} (j_1-j_2+w) \right]
    & \!\!\text{, \cite{Alev:EtAl:2020}, } \\[3.2ex]
    \displaystyle\frac{1}{w^2} \left[ 1 + \frac{2}{w^2} \sum_{j=1}^{w-1} j^2 \right]
    = \frac{1}{w^2} \left[ 1 + \frac{(w-1)(2w-1)}{3w} \right]
    & \!\!\text{, summing squared weights.}
\end{cases}
\end{align*}
\end{small}
Some tedious algebra shows that the two last two expressions are equivalent. \cite{Alev:EtAl:2020} provided a few comparisons
between MA, DMA, EWMA and CUSUM charts. Unfortunately, the opponents
are chosen in either an unfair (take the same $w$ for MA and DMA charts) or in an arbitrary
way with EWMA and CUSUM charts. Here, we match MA($w_1$), DMA($w_2$)
and EWMA($\lambda$), aiming at roughly the same final variance of the plotted statistic.
Thus, starting with some $w_2$ for the DMA chart, we determine
\begin{equation*}
  \sigma_D^2 = \frac{1}{w_2^2} \left[ 1 + \frac{(w_2-1)(2w_2-1)}{3w_2} \right] \,,\quad
  w_1 = \left[ \frac{1}{\sigma_D^2} \right] \,,\quad
  \lambda = \frac{2\sigma_D^2}{\sigma_D^2 + 1}\,.
\end{equation*}
Here $[\cdot]$ denotes rounding to the next integer.

Now we re-arrange some zero-state ARL results from Tables 3 and 4
in \cite{Alev:EtAl:2020} in blocks in Table~\ref{tab:dmaARL}.
\begin{table}[hbt]
\centering
\caption{Zero-state ARL of DMA($w_2$), MA($w_1$), and EWMA($\lambda$); nominal in-control
ARL of 370.}\label{tab:dmaARL}
\small
\begin{tabular}{cccccccccccc} \toprule
  & \multicolumn{10}{c}{$\delta$} \\[0.3ex]
  & 0 & 0.2 & 0.4 & 0.6 & 0.8 & 1.0 & 1.25 & 1.5 & 2 & 3 \\ \midrule
  $w_2=2$ &
  371.5 & 250.2 & 119.4 & 57.7 & 29.7 & 17.4 & 9.7 & 6.3 & 3.4 & 1.7 \\ 
  $w_1=3$ &
  369.8 & \B 244.7 & \B 109.4 & \B 51.1 & \B 26.4 & \B 15.4 & \B 8.8 & \B 5.7 & \B 3.1 & \B 1.6 \\ 
  $\lambda=0.545$ &
  370.0 & 246.0 & 114.0 & 54.0 & 28.3 & 16.4 & 9.4 & 6.1 & 3.3 & 1.6 \\ \midrule
  $w_2=3$ &
  370.3 & \B 218.0 & \B 90.5 & \B 40.8 & 21.6 & 12.9 & 7.7 & 5.4 & 3.3 & 1.7 \\ 
  $w_1=4$ &
  370.5 & 222.4 &  92.0 & 42.3 & \B 21.5 & \B 12.6 & \B 7.3 & \B 4.9 & \B 2.8 & \B 1.6 \\ 
  $\lambda=0.380$ &
  370.0 & 212.0 &  84.5 & 37.8 & 19.8 & 11.9 & 7.3 & 5.0 & 3.0 & 1.6 \\ \midrule
  $w_2=4$ &
  369.6 & 198.9 &  75.3 & 33.6 & 18.1 & 10.9 & 7.1 & 5.1 & 3.3 & 1.6 \\ 
  $w_1=6$ &
  370.8 & \B 193.8 & \B 71.8 & \B 31.4 & \B 16.5 & \B 10.0 & \B 6.3 & \B 4.4 & \B 2.7 & \B 1.5 \\ 
  $\lambda=0.293$ &
  370.0 & 189.7 &  69.5 & 30.6 & 16.4 & 10.2 & 6.5 & 4.6 & 2.8 & 1.6 \\ \midrule
  $w_2=5$ &
  369.7 & 186.0 & \B 65.5 & 29.6 & 15.9 & 10.2 & 7.0 & 5.2 & 3.3 & 1.6 \\ 
  $w_1=7$ &
  370.8 & \B 183.2 & \B 65.5 & \B 28.6 & \B 15.2 & \B 9.4 & \B 6.0 & \B 4.3 & \B 2.7 & \B 1.5 \\ 
  $\lambda=0.239$ &
  370.0 & 173.6 &  60.5 & 26.7 & 14.6 &  9.3 & 6.1 & 4.4 & 2.8 & 1.6 \\ \midrule
  $w_2=6$ &
  370.6 & 172.4 & 60.0 & 26.8 & 15.0 & 9.9 & 6.9 & 5.2 & 3.2 & \B 1.5 \\
  $w_1=9$ &
  370.4 & \B 166.7 & \B 56.6 & \B 24.8 & \B 13.6 & \B 8.6 & \B 5.7 & \B 4.1 & \B 2.6 & \B 1.5 \\
  $\lambda=0.202$ &
  370.0 & 161.4 &  54.4 & 24.3 & 13.6 &  8.8 & 5.9 & 4.3 & 2.7 & 1.5 \\ \bottomrule
\end{tabular}
\end{table}
Additional values (for $w_2=6$ and
for $w_1 \in \{6, 7, 9\}$) were estimated with Monte-Carlo simulations ($10^7$ replications).
The EWMA chart values were
determined with the \textsf{R} package \texttt{spc}.
To make comparisons between the MA and DMA charts easy (EWMA is left out for the moment), we bolded the smaller ARL per block and $\delta$. From
the results we conclude that
the simpler and older MA chart performs for the most part better than the DMA chart.

\begin{figure}[hbt]
\centering
\renewcommand{\tabcolsep}{0mm}
\begin{tabular}{cc} 
  \tiny $\delta = 0.6$ & \tiny $\delta = 1$ \\[-1ex]
  \includegraphics[width=.5\textwidth]{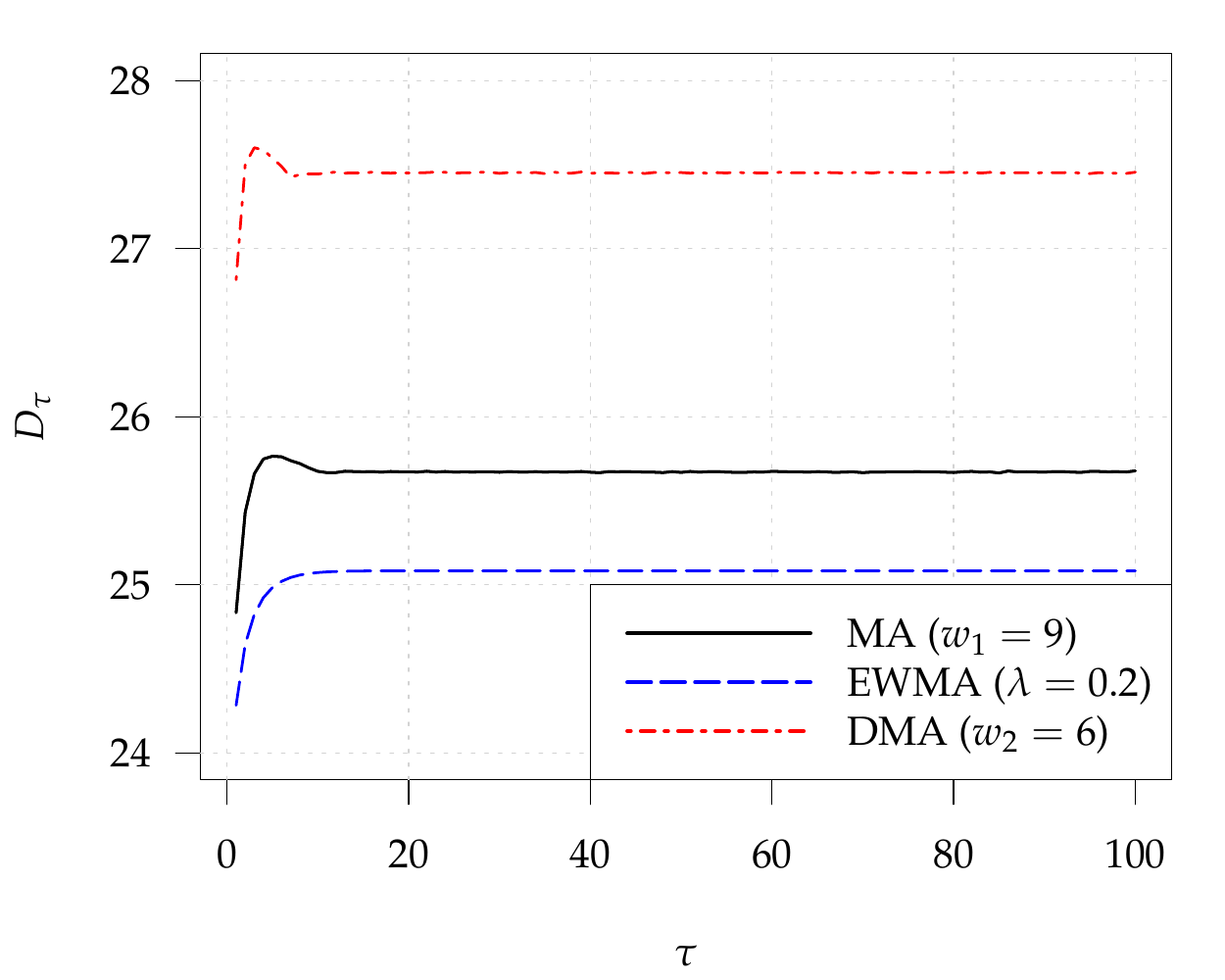} & \includegraphics[width=.5\textwidth]{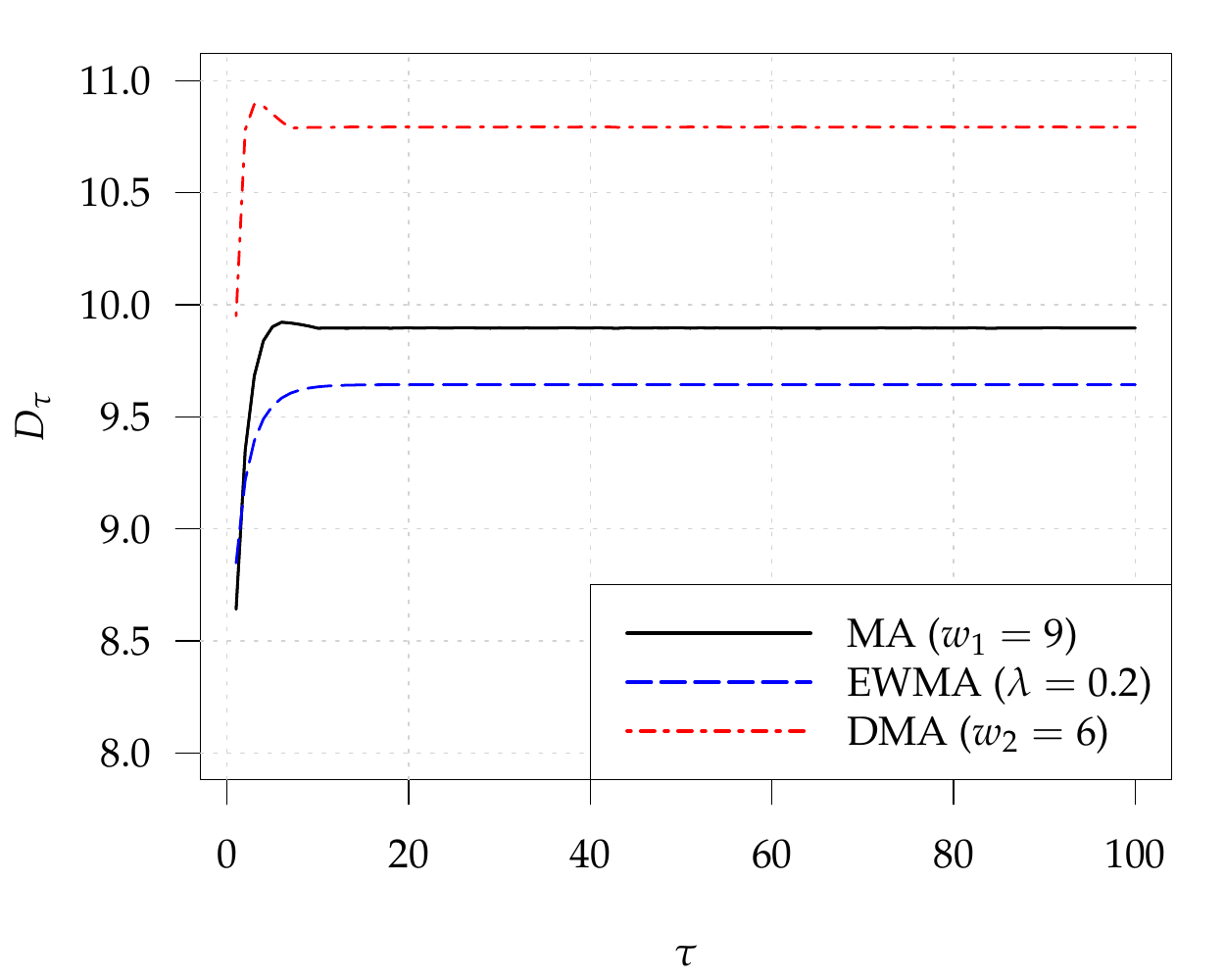} \\[-1ex]
  \tiny $\delta = 2$ & \tiny $\delta = 3$ \\[-1ex]
  \includegraphics[width=.5\textwidth]{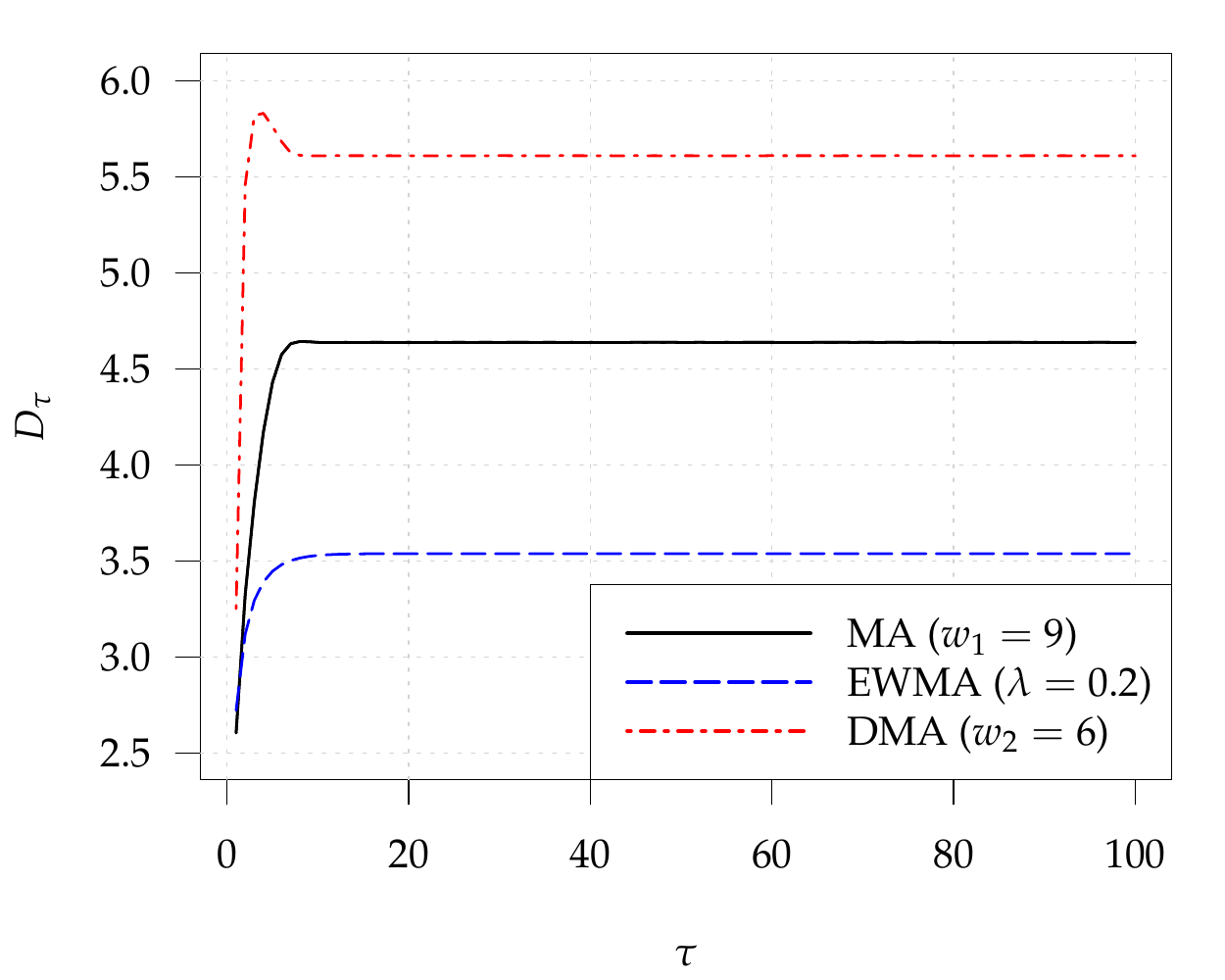} & \includegraphics[width=.5\textwidth]{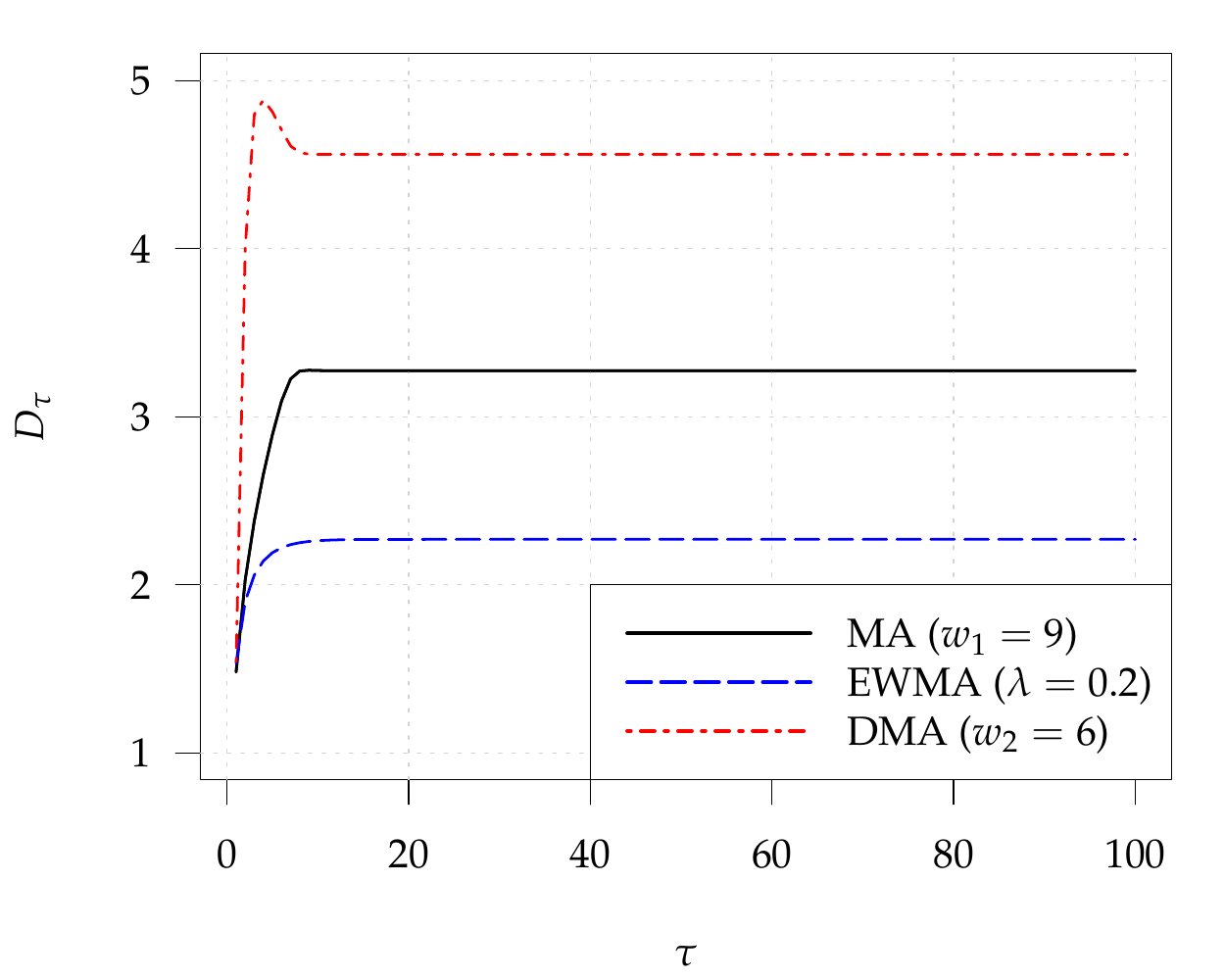}
\end{tabular}
\caption{CED profiles of DMA($w_2$), MA($w_1$), and EWMA($\lambda$), i.\,e.
$D_\tau = E_\tau(L-\tau+1\mid L\ge \tau)$ vs. change-point position $\tau$, 
for four shifts $\delta$; in-control (zero-state) ARL is $A = 370$.}
\label{fig:madma_little}
\end{figure}

\cite{Wong:Gan:Chan:2004} stated:
\textit{``For zero state [ARL], it was found that among all MA charts
(with different values of $w$) with the same in-control ARL,
the chart with the largest $w$ is always the quickest in detecting any finite shift.
In other words, no optimal chart within the class of MA charts can be found.''}
Similar patterns could be observed for the DMA chart
\citep[slight deviations in Table 3 of][are probably Monte Carlo artefacts]{Alev:EtAl:2020}.
To avoid arbitrariness, one has to move away from the zero-state ARL
competition. For instance, \cite{Wong:Gan:Chan:2004} followed
an old design principle, \textit{``the steady state run length will be considered''},
which we apply here as well considering the CED. Here, we pick the configurations
from the last block of Table~\ref{tab:dmaARL}
and determine $\{D_\tau\}_{\tau=1}^{100}$, see Figure~\ref{fig:madma_little}.
The results for the MA and DMA charts were obtained again by Monte Carlo simulations
($10^7$ replications), whereas the EWMA values were calculated with
\texttt{xewma.arl()} from the \textsf{R} package \texttt{spc}.
The CED profiles of the three control charts mirror the ordering of
the corresponding zero-state ARL results. We can briefly summarize the four
diagrams in Figure~\ref{fig:madma_little} with two
statements. It is easy to find a MA control chart which performs overall
better than any given DMA control chart. Second, the steady-state ARL
of the DMA chart could be substantially larger than the zero-state ARL
(see $\delta = 2$ and $= 3$).

\begin{figure}[hbt]
\centering
\renewcommand{\tabcolsep}{0mm}
\begin{tabular}{cc} 
  \small MA & \small DMA \\[-1ex]
  \includegraphics[width=.51\textwidth]{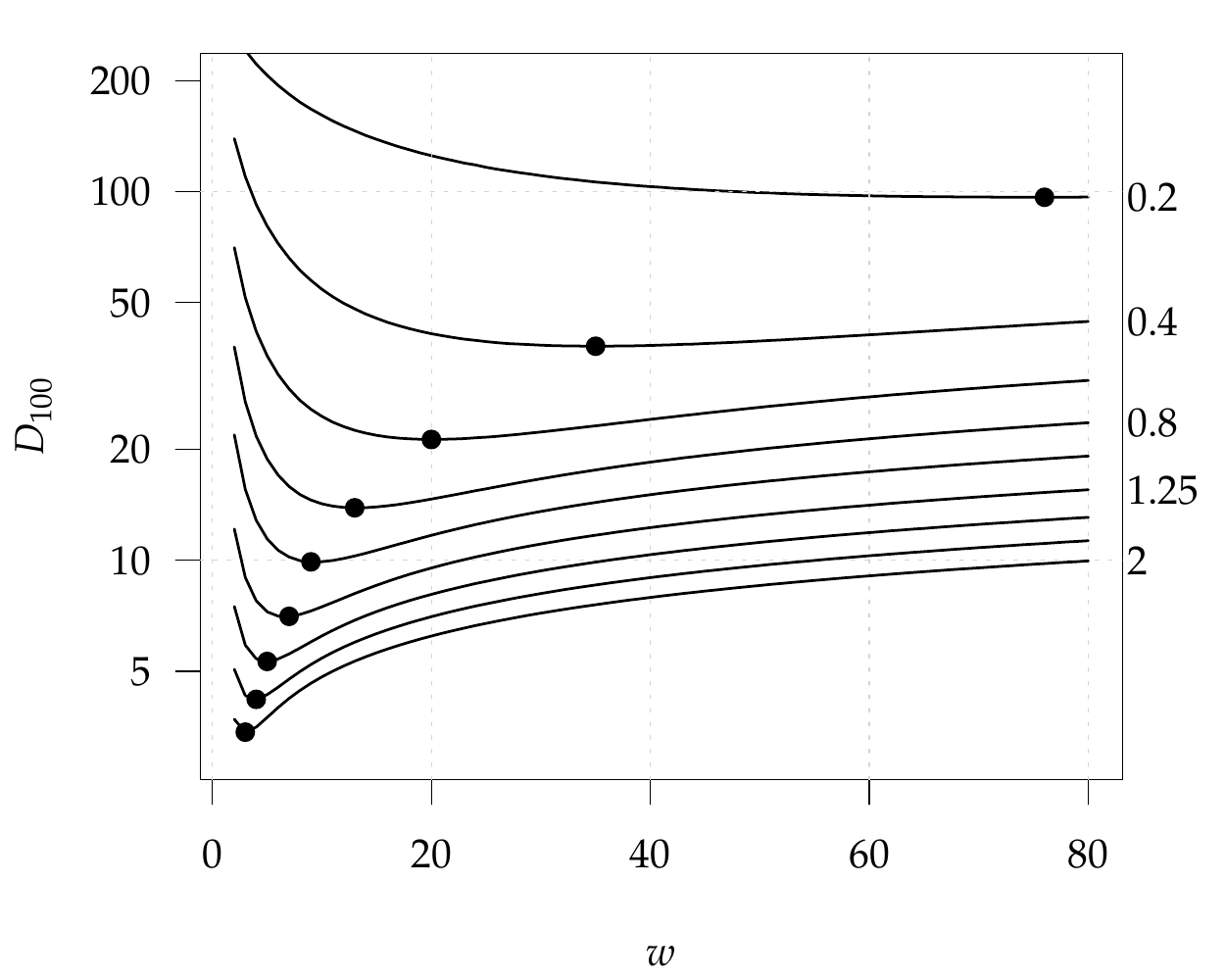} & \includegraphics[width=.51\textwidth]{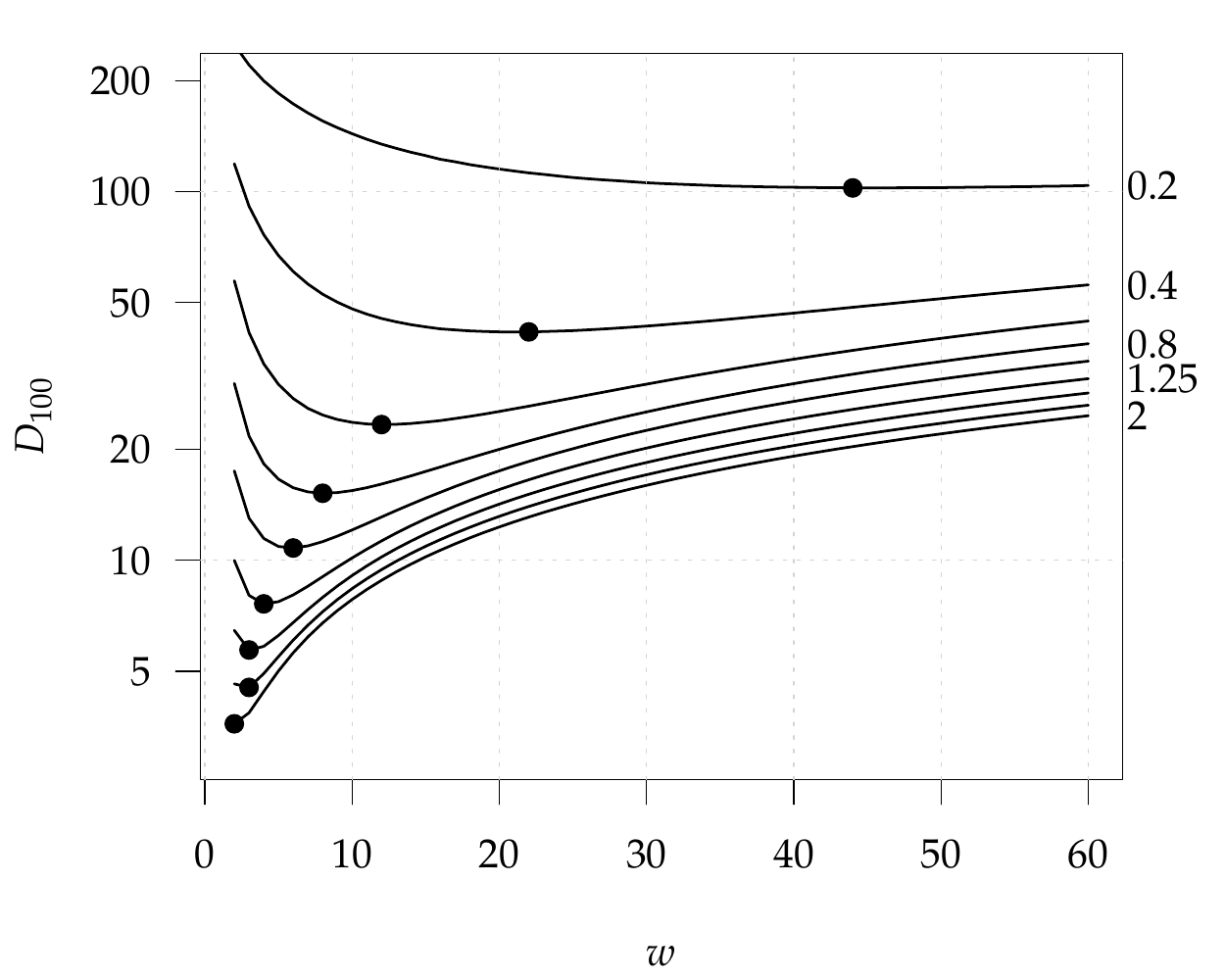}
\end{tabular}
\caption{$D_{100}$ profiles of MA and DMA vs. $w$ for
various shifts $\delta \in \{0.2, 0.4, \ldots, 2\}$,
optimal choice $w$ per shift marked with $\bullet$;
 in-control (zero-state) ARL is $A = 370$.}
\label{fig:madma_optim}
\end{figure}

\begin{figure}[hbt]
\centering
\renewcommand{\tabcolsep}{0mm}
\begin{tabular}{cc} 
  \small $\delta=0.6$ optimal & \small $\delta=1.5$ optimal \\[-1ex]
  \includegraphics[width=.51\textwidth]{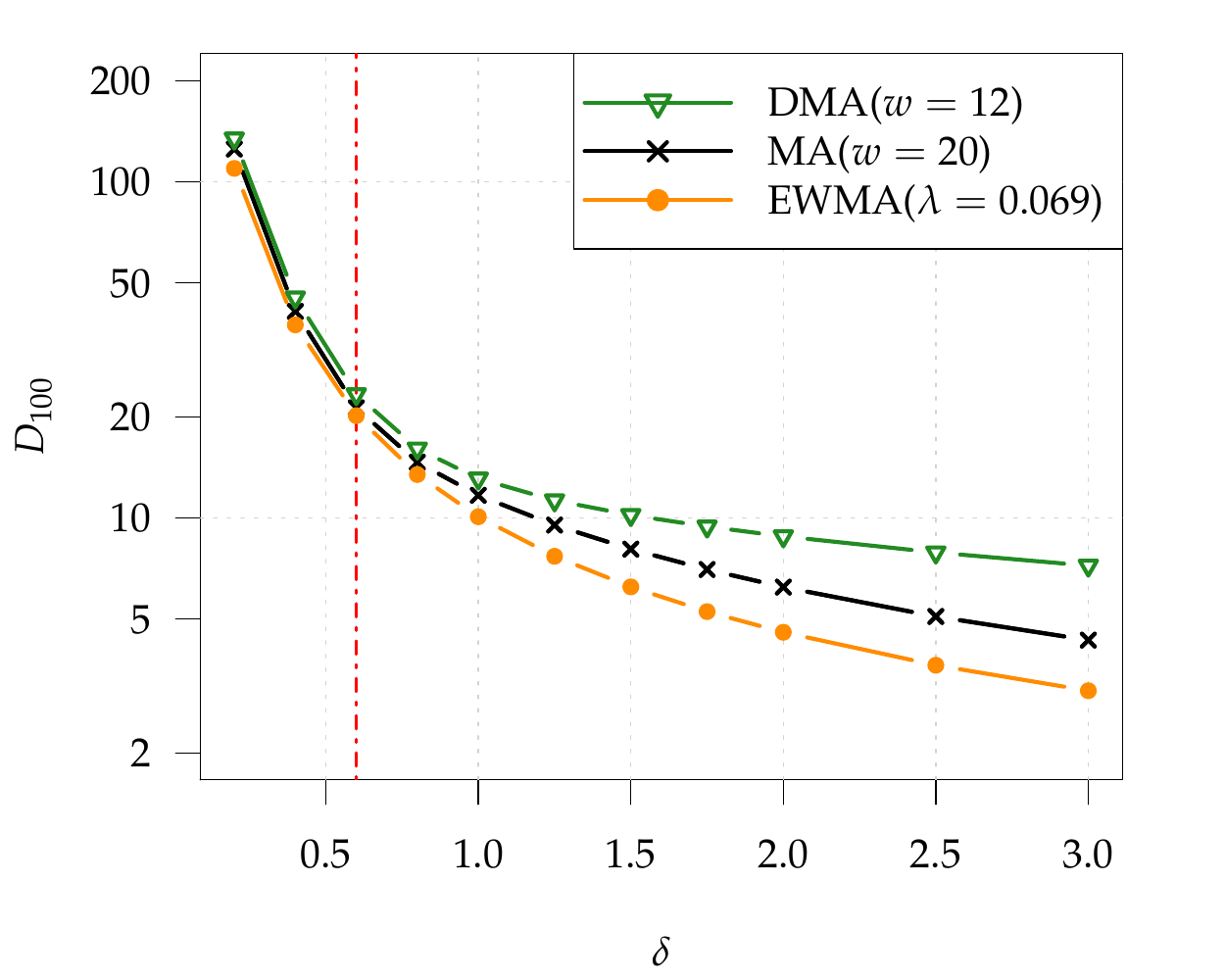} & \includegraphics[width=.51\textwidth]{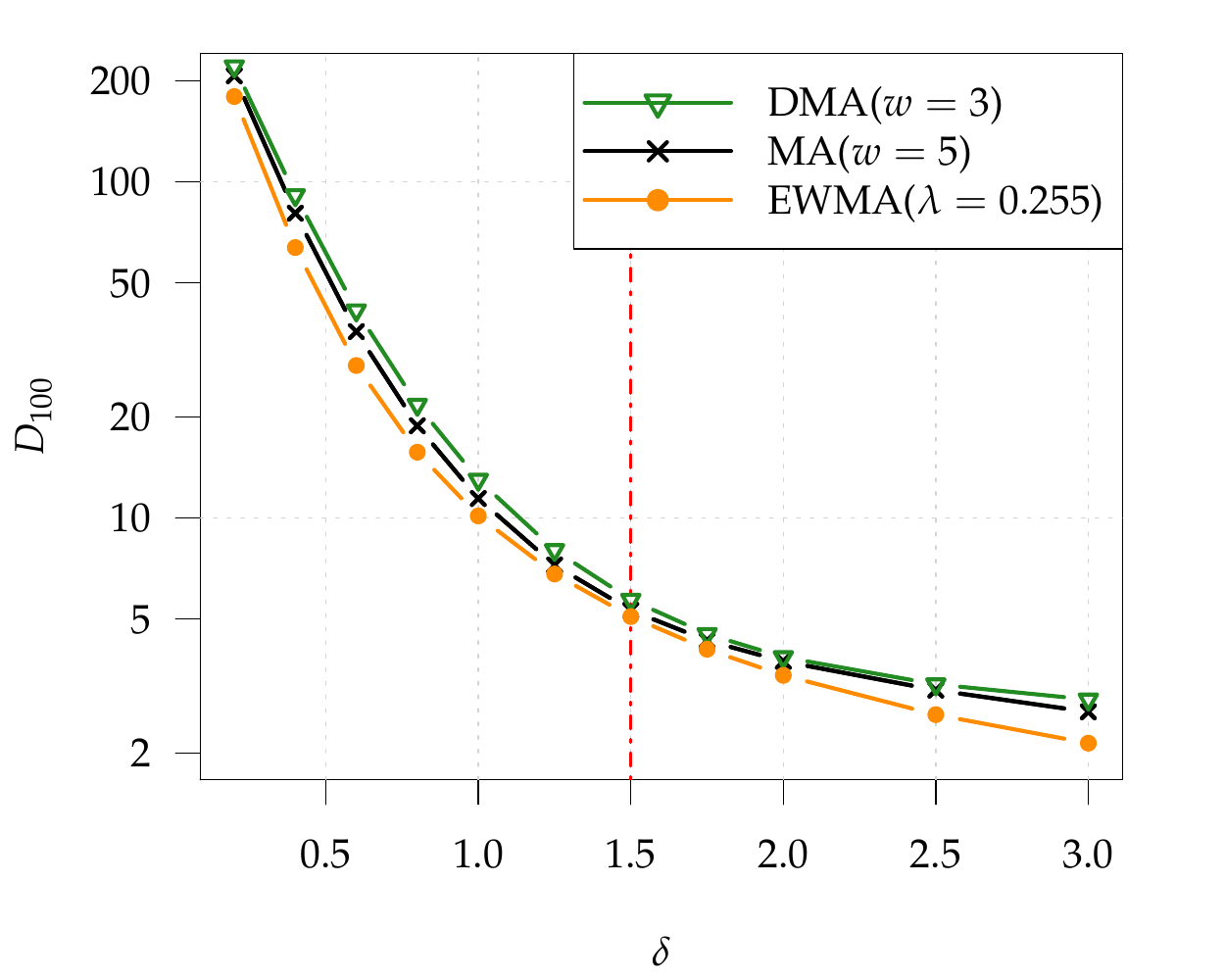}
\end{tabular}
\caption{Steady-state ARL ($D_{100}$) profiles of MA,  DMA and EWMA vs. $\delta$;
in-control (zero-state) ARL is $A = 370$.}
\label{fig:madmaewma_sARL}
\end{figure}

To complete this steady-state ARL comparison, we utilize $D_{100}$ as
proxy (motivated by the stable convergence patterns in Figure~\ref{fig:madma_little}) and conduct an investigation 
concerning setups minimizing the steady-state ARL for a given shift,
the results are given in Figure~\ref{fig:madma_optim}.
For the MA and DMA charts, we picked values $w$ up to 80 and 60, respectively,
and estimated the CED at $\tau=100$, namely $D_{100}$, by
using a Monte Carlo simulations with $10^7$ replications. From the resulting
profiles ($D_{100}$ vs. $w$) we observe a couple of facts.
Only for $\delta=0.2$ (which is a more of an academic choice) and to lesser extent for $\delta = 0.4$, the optimum $w$ values are substantially large.
Second, the DMA curves increase more profoundly than for the MA chart
in case of medium and large shifts. This signals trouble for these shifts if one
aims at optimizing the DMA chart for small shifts.
It should be noted that all MA minima are smaller than their DMA
counterparts. Thus, if one is interested in minimizing
the steady-state ARL at a single shift $\delta$, then the MA chart will beat the DMA chart
for all settings we considered.
The optimal choices of $w$ for the MA and DMA charts are marked and collected in the
Table~\ref{tab:optW} together with the respective
$\lambda$ values of the EWMA chart. From these configurations, we select
the ones for $\delta = 0.6$ and $\delta = 1.5$ to
demonstrate the steady-state ARL behavior over a range of shifts.
\begin{table}[hbt]
\centering
\caption{Optimal (to minimize $D_{100}$) choices of $w$ and $\lambda$ for MA, DMA and EMWA; in-control ARL is 370.}\label{tab:optW}
\begin{tabular}{cc*{11}{c}} \toprule
 $\delta$  && 0.2 & 0.4 & 0.6 & 0.8 & 1 & 1.25 & 1.5 & 1.75 & 2 & 2.5 & 3 \\ \midrule
 $w_1$     &&  76 &  35 &  20 &  13 & 9 &    7 &   5 &    4 & 3 &   2 & 2 \\
 $w_2$     &&  44 &  22 &  12 &   8 & 6 &    4 &   3 &    3 & 2 &   2 & 2 \\ 
 $\lambda$ && 0.02 & 0.04 & 0.07 & 0.10 & 0.14 & 0.20 & 0.26 & 0.32 & 0.38 & 0.53 & 0.69 \\ \bottomrule
\end{tabular}
\end{table}
The two EWMA values, $0.07$ and $0.26$, are quite common choices (note that for
all calculations we used three digits, that is, the actual
$\lambda = 0.069$ and $= 0.255$). The same is true
for the MA chart \citep[$w_1 = 20$ and $= 5$][considered values up to $w_1 = 30$]{Wong:Gan:Chan:2004} and DMA \citep[$w_2 = 12$ and $= 4$][dealt with values up to $w_2 = 15$]{Wong:Gan:Chan:2004}. For these two settings, we determined
the steady-state ARL (again using $D_{100}$ and Monte Carlo simulation, except for the EWMA chart).
The resulting profiles are given in Figure~\ref{fig:madmaewma_sARL}.
The three schemes are clearly ordered. For all shifts $\delta$, the EWMA chart is the best and the MA chart is the second best scheme. In particular for the $\delta = 0.6$ designs, the three curves clearly  begin to diverge for increasing values of $\delta$.
Given the specific weighting patterns of the DMA chart, this is not surprising.

In summary, the DMA control chart design should not be used.
This statement is valid as well for the even more complex
triple MA \citep{Alev:Chat:Kouk:2021e} and quadruple MA \citep{Alev:Chat:Kouk:2021d} charts. Note that the computation
of the variance of the respective statistics is a voluminous task
resulting in various theorems in \cite{Alev:Chat:Kouk:2021e, Alev:Chat:Kouk:2021d}.
That is, we obtain complex charts with demanding setup procedures and
cumbersome Monte Carlo studies to evaluate these schemes. All three charts, DMA, TMA and QMA, exhibit the
undesirable weighting pattern that the highest weight is set into the center of their respective
windows, whereas the most recent data values get very low weights. This delays the
detection of changes that happen later than instantly at the startup.

\section{DEWMA and TEWMA Charts} \label{sec:dewma}

The first use of two EWMA statistic iterations were proposed by
\cite{Sham:Amin:Sham:1991} and \cite{Sham:Sham:1992}.
However, they did not made such an impact as the later equivalent proposals by \cite{Zhan:Govi:Lai:Bebb:2003} and \cite{Zhan:Chen:2005}.
The more recent papers focused on the simpler case, where the smoothing
constants of both EWMA series are the same, whereas the previous
publications dealt with the more general case.
Despite the critical statements regarding double EWMA (DEWMA) control charts
in \cite{Mahm:Wood:2010}, DEWMA charts gained quite some popularity
in the SPM literature. Therefore it is time
to clarify in a different way that
DEWMA charts are inferior to the simpler EWMA charts.

To begin, we introduce some notation. Note that we deal with the
more popular case, where both EWMA equations are based on the same $\lambda$.
Then, writing $\bar{\lambda} := 1-\lambda$, we  have the following equations:
\begin{align*}
  Z_0^{(1)} & = \mu_0 \,,\; Z_0^{(2)} = \mu_0  \,. \\
  Z_i^{(1)} & = (1-\lambda) Z_{i-1}^{(1)} + \lambda X_i \,, \\
  Z_i^{(2)} & = (1-\lambda) Z_{i-1}^{(2)} + \lambda Z_i^{(1)} \\
  & = \lambda^2 \sum_{j=1}^i (i-j+1)\bar{\lambda}^{i-j} X_j + i\lambda\bar{\lambda}^i Z_0^{(1)} + \bar{\lambda}^i Z_0^{(2)} \,.
\end{align*}
From the above publications, we obtain for the variance of $Z_t^{(2)}$
\begin{align}
  \sigma_i^2 & = \lambda^4
  \frac{1 + \bar{\lambda}^2 - (i^2+2i+1)\bar{\lambda}^{2i} + (2i^2 + 2i -1) \bar{\lambda}^{2i+2} - i^2 \bar{\lambda}^{2i+4}}{(1 - \bar{\lambda}^2)^3}
  \,,  \label{eq:dewmaVt} \\
  \sigma_\infty^2 & = \lambda^4 \frac{1 + \bar{\lambda}^2}{(1 - \bar{\lambda}^2)^3} \,. \label{eq:dewmaVinf}
\end{align}
Then by using \eqref{eq:dewmaVt},
the run-length of the DWMA chart is
\begin{equation*}
  L_{DE} =
  \min \left\{ i\ge 1\!: |Z_i^{(2)} - \mu_0| > c_{DE} \sigma_i \,\right\} \,.
\end{equation*}
Obviously, setup, deployment and analysis of the DEWMA chart is more
complicated than for the EWMA chart. Monte Carlo analyses are required.
For the EWMA chart and DEWMA chart competition, \cite{Zhan:Chen:2005} (and many others)
picked the same $\lambda$ for the EWMA chart as for the DEWMA chart. This rather oversimplified and inappropriate
approach is omnipresent in the literature, unfortunately.
\cite{Mahm:Wood:2010} went a different way. They proposed to set the EWMA chart
$\lambda$ in such a way that for a given DEWMA design the weighting
patterns of both competing charts feature the same maximum value.
Here, we again aim at the same asymptotic variance, that is, choose
$\lambda_E$ so that $\lambda_E/(2-\lambda_E)$ is equal to
the respective DEWMA value in \eqref{eq:dewmaVinf}.
For example, starting with DEWMA\,($\lambda=0.1$) we get (roughly) to
EWMA\,($\lambda=0.05$). 
\cite{Mahm:Wood:2010} determined the smaller $0.03874$ 
EWMA smoothing constant. It turns out that aiming
at equal asymptotic variances leads to larger values of $\lambda$
than aiming at equal maximum weights. We will see in a moment that
the slightly smaller change yields competitive EWMA
designs with respect to the zero-state ARL.

Next, we present in Table~\ref{tab:dewma_zARL} some results of
\begin{table}[hbt]
\centering
\caption{Fragments of Table 4 in \cite{Zhan:Chen:2005} and
some more accurate numbers (labelled $10^8$ and \texttt{spc});
zero-state ARL of DEWMA (upper block) and EWMA (lower block); in-control ARL 200.}\label{tab:dewma_zARL}
\small
\begin{tabular}{ccccccccccl} \toprule
  & \multicolumn{9}{c}{shift $\delta / \sqrt{5}$} & \\
  & 0 & 0.1 & 0.2 & 0.3 & 0.4 & 0.5 & 1 & 1.5 & 2 & \\ \midrule
  ZC2005 & 200.0 & 57.5 & 20.1 & 10.7 & 6.8 & 4.7 & 1.6 & 1.1 & 1.0 & $\lambda=0.1$ \\
  $10^8$ & 199.9 & 57.3 & 20.4 & 10.7 & 6.7 & 4.7 & 1.6 & 1.1 & 1.0 & $\lambda=0.1$ \\ \midrule
  %
  ZC2005 & 200.1 & 69.2 & 24.0 & 12.4 & 7.8 & 5.4 & 1.9 & 1.2 & 1.0 & $\lambda=0.1$ \\
  ZC2005 & 200.2 & 57.5 & 20.2 & 10.7 & 6.9 & 4.8 & 1.7 & 1.1 & 1.0 & $\lambda=0.05$ \\
  \texttt{spc} & 200.0 & 56.9 & 20.4 & 10.7 & 6.8 & 4.8 & 1.7 & 1.1 & 1.0 & $\lambda=0.05$ \\ \bottomrule
\end{tabular}
\end{table}
Table 4 in \cite{Zhan:Chen:2005} for an in-control ARL value of 200.
\cite{Zhan:Chen:2005} compared the $\lambda=0.1$ results, for example,
and concluded
\textit{``The DEWMA mean chart performs better than the EWMA mean chart when the process mean shifts are smaller than one half of the process standard deviation''}. Of course, EWMA\,($\lambda=0.1$) exhibits larger zero-state
ARL values for roughly all considered shifts. Doing the comparison, however, more
appropriately, we compare with the EWMA chart with \,$\lambda=0.05$ and see from Table~\ref{tab:dewma_zARL}
that the zero-state ARL values are roughly the same as the DEWMA chart values in the top part of the table.
It remains unclear, why \cite{Zhan:Chen:2005} did not spot this fact. 

\begin{figure}[hbt]
\centering
\renewcommand{\tabcolsep}{0mm}
\begin{tabular}{cc} 
  \small $\delta = 0.2\sqrt{5}$ & \small $\delta = 0.5\sqrt{5}$ \\[-1ex]
  \includegraphics[width=.5\textwidth]{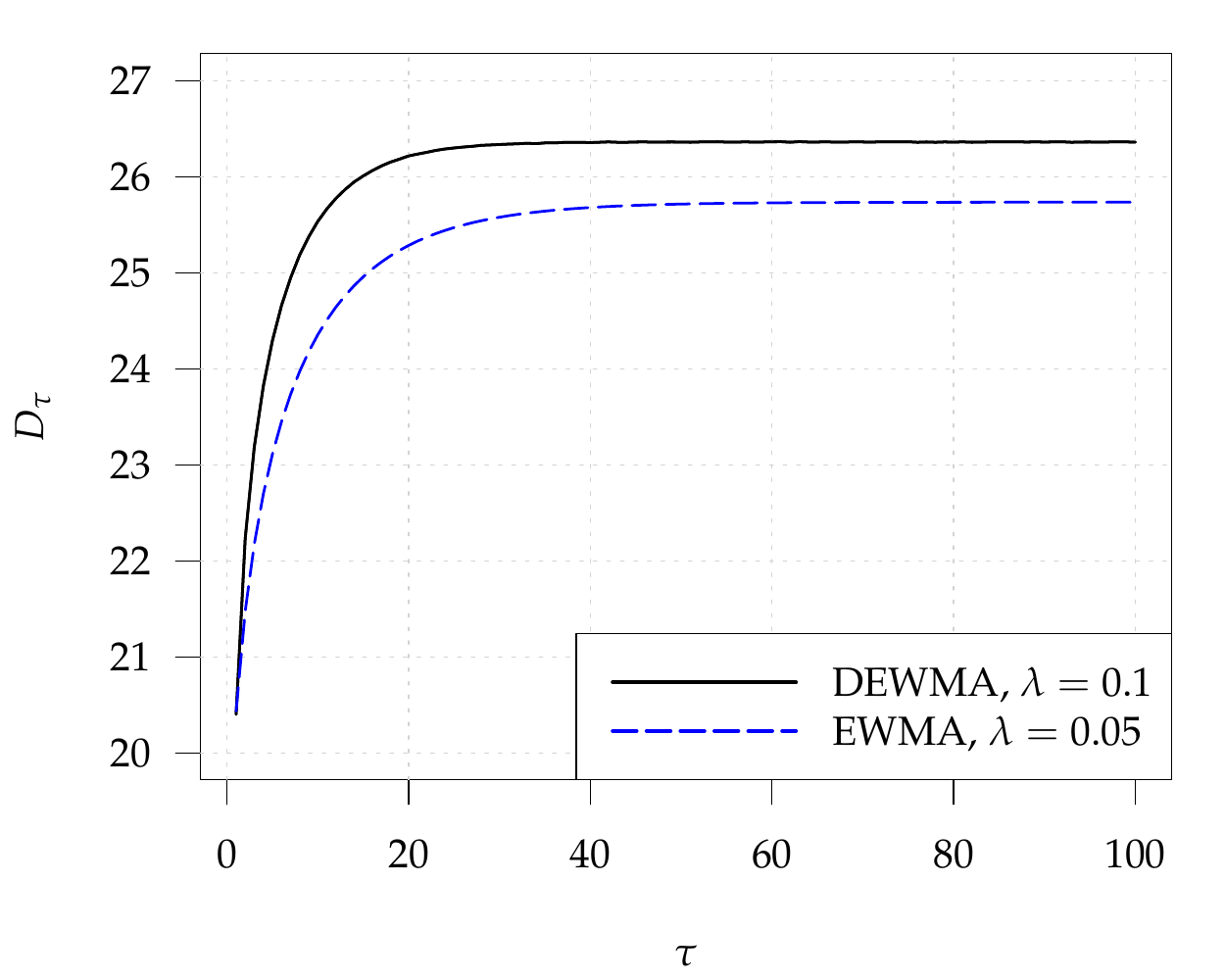} & \includegraphics[width=.5\textwidth]{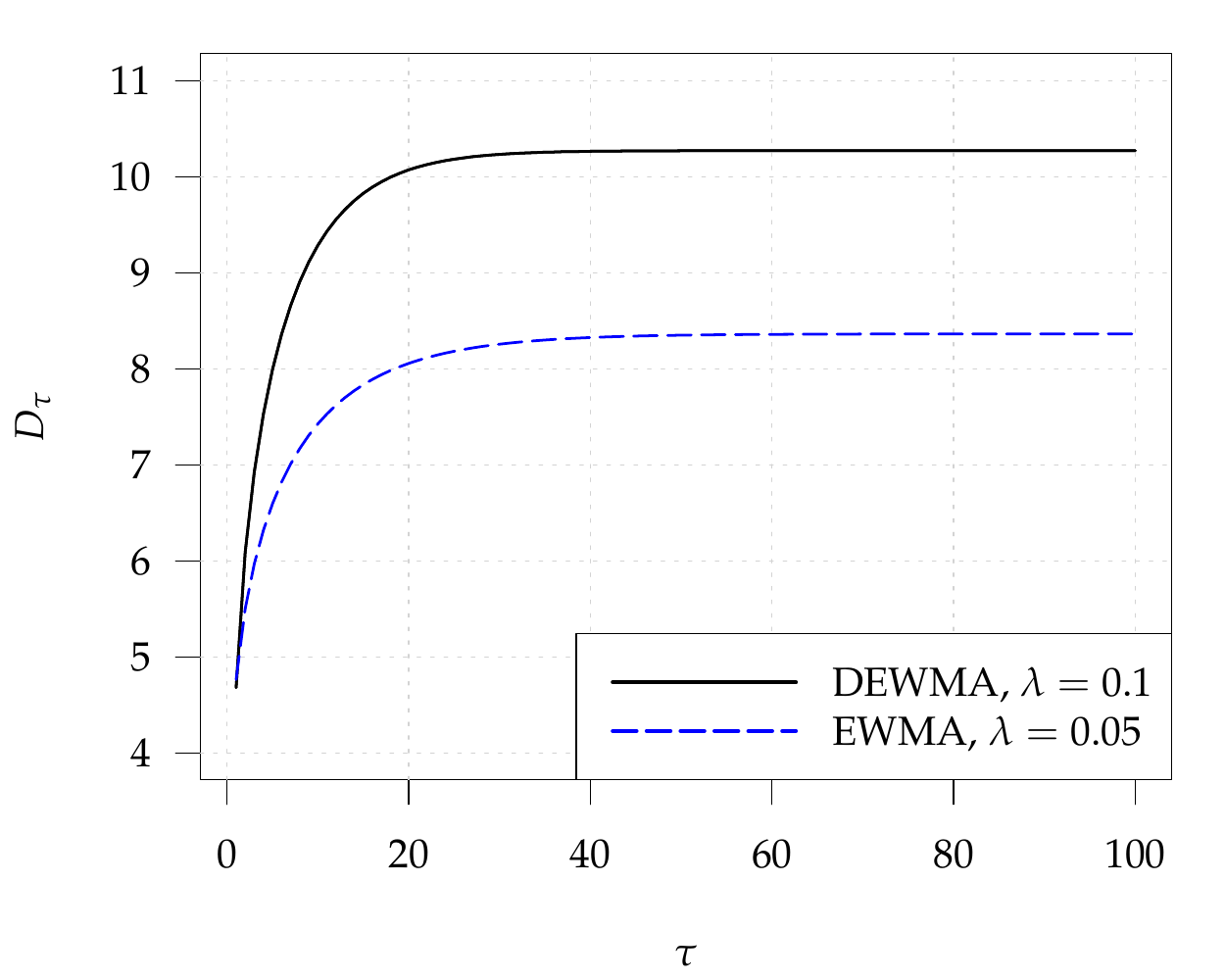} \\[-1ex]
  \small $\delta = 1\sqrt{5}$ & \small $\delta = 2\sqrt{5}$ \\[-1ex]
  \includegraphics[width=.5\textwidth]{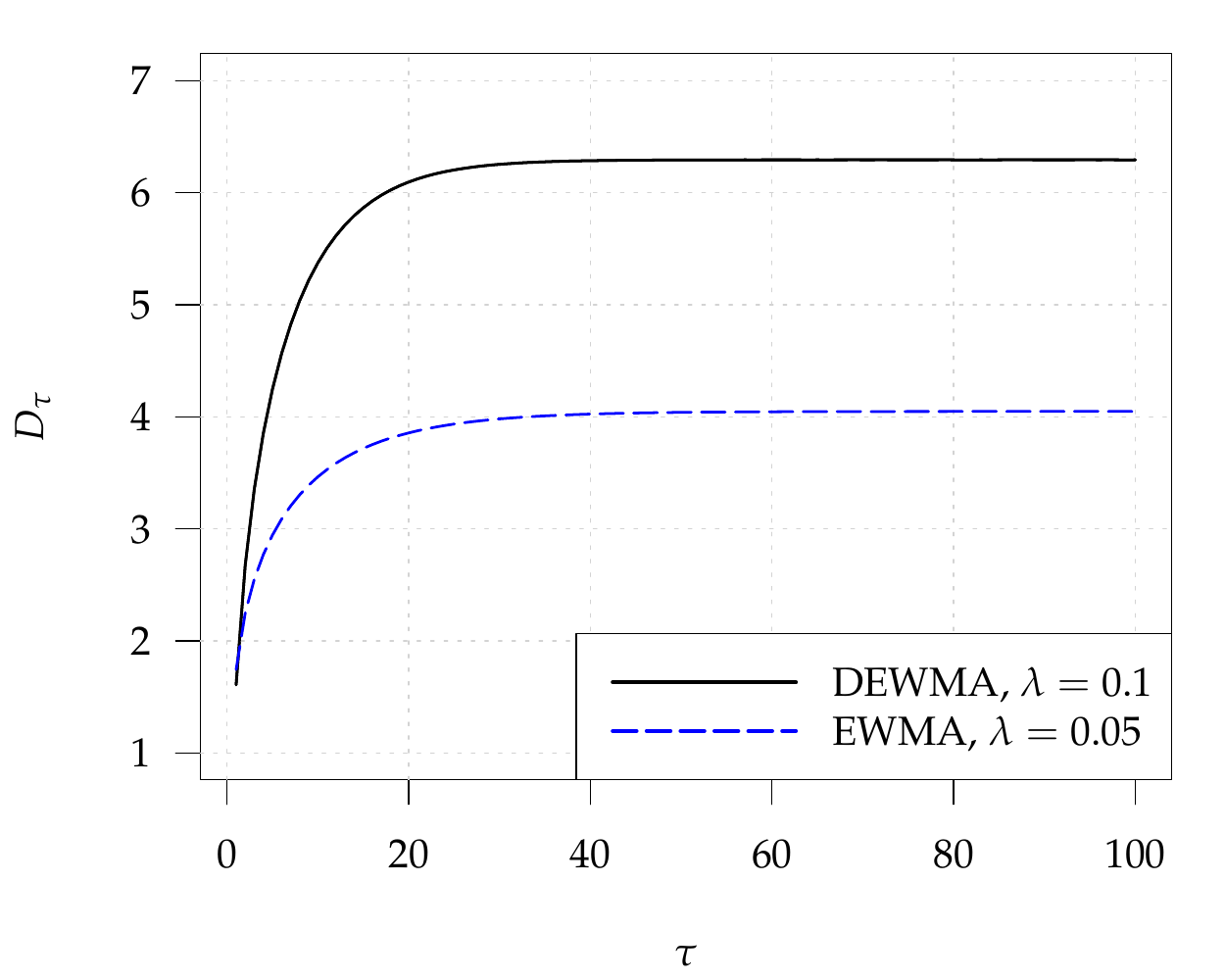} & \includegraphics[width=.5\textwidth]{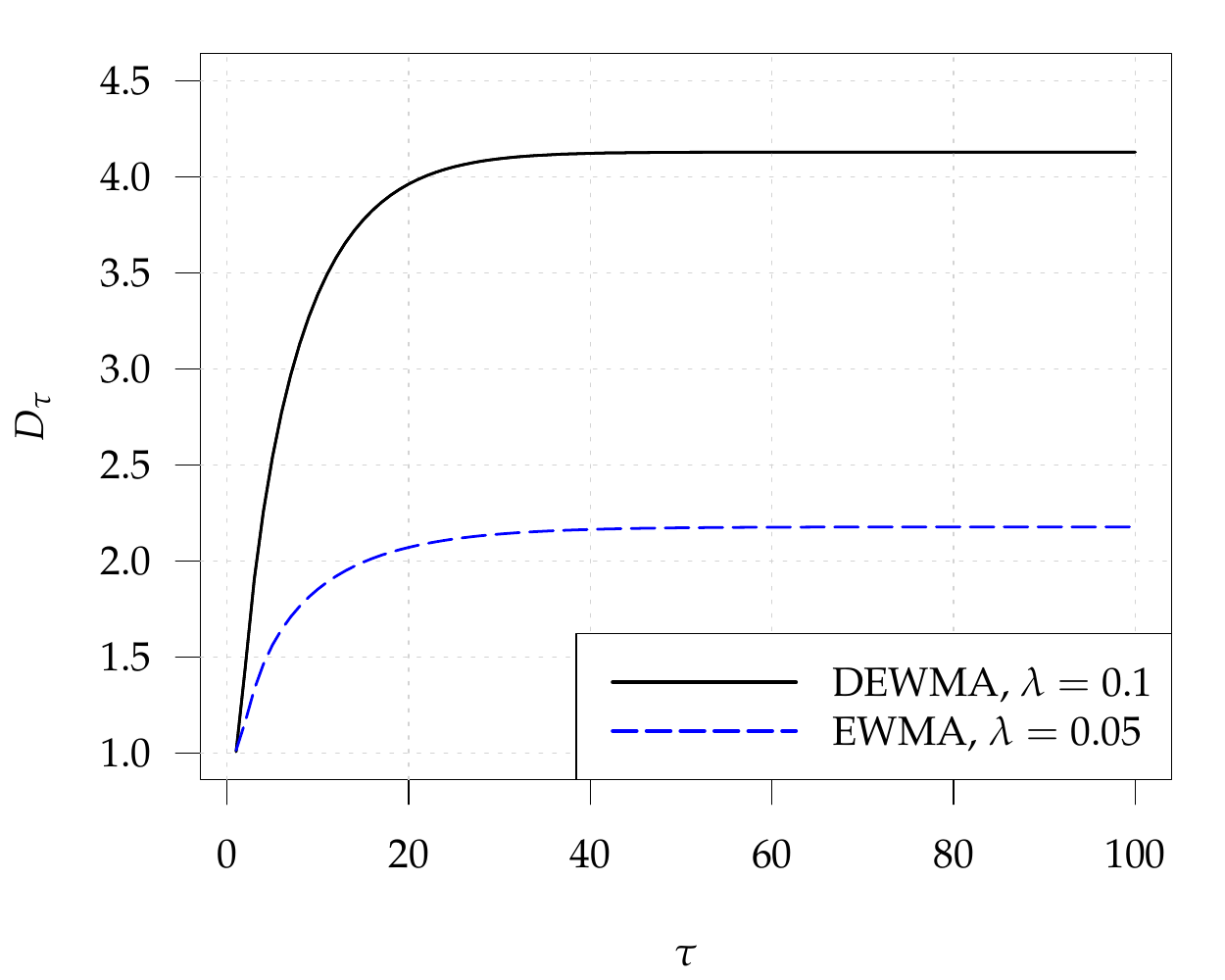}
\end{tabular}
\caption{CED profiles of DEWMA\,($\lambda=0.1$) and EWMA\,($\lambda=0.05$), i.\,e.
$D_\tau = E_\tau(L-\tau+1\mid L\ge \tau)$ vs. change-point position $\tau$, 
for four shifts $\delta$; in-control (zero-state) ARL is $A = 200$.}
\label{fig:dewma_CED}
\end{figure}

Knowing that further decreasing the value of $\lambda$ we would
achieve even better zero-state ARL values of the EWMA chart for all shifts,
we turn to more important properties such as CED and
conditional steady-state ARL. Taking again the pair
DEWMA\,($\lambda=0.1$) and EWMA\,($\lambda=0.05$),
we calculated the CED for all instances in Table~\ref{tab:dewma_zARL}
and provide the resulting curves for four shifts in
Figure~\ref{fig:dewma_CED}.
From the curves in Figure~\ref{fig:dewma_CED} we learn that
the zero-state ($\tau=1$) values are roughly equal as we know
from Table~\ref{tab:dewma_zARL}, whereas for later changes
the EWMA chart always exhibits lower CED values. The larger the shift,
the more pronounced is the difference. For both control charts,
the CED series stabilizes sufficiently quick so that we
could use $D_{100}$ as the conditional steady-state ARL and
as representative delay for changes after $\tau=30$. Next we
present the resulting values in Table~\ref{tab:dewma_sARL}.
Note that we added results for triple EWMA (TEWMA) following \cite{Alev:Chat:Kouk:2021c},
all calculated with $10^8$ replications, smoothing
constant $\lambda = 0.13$ (to achieve the same asymptotic variance
as DEWMA with $\lambda = 0.1$), and a control limit factor
$L = 1.91$ (in-control ARL 200). For details regarding the exact variance etc., we refer to \cite{Alev:Chat:Kouk:2021c}.
\begin{table}[hbt]
\centering
\caption{Augmenting Table 4 in \cite{Zhan:Chen:2005}
with some steady-state ARL results, for DEWMA with Monte Carlo
simulation ($10^8$ replicates) and for EWMA with the \textsf{R}
package \texttt{spc}; in-control ARL 200; TEWMA ($10^7$ replicates) numbers are added too.}\label{tab:dewma_sARL}
\small
\begin{tabular}{rcccccccccl} \toprule
  & \multicolumn{9}{c}{shift $\delta / \sqrt{5}$} & \\
  & 0 & 0.1 & 0.2 & 0.3 & 0.4 & 0.5 & 1 & 1.5 & 2 & \\ \midrule
  TEWMA & -- & 67.4 & 27.9 & 18.3 & 14.6 & 12.6 & 8.5 & 7.0 & 6.1 & $\lambda=0.13$ \\
  DEWMA & -- & 66.1 & 26.4 & 16.4 & 12.4 & 10.3 & 6.3 & 4.9 & 4.1 & $\lambda=0.1$ \\
   EWMA & -- & 64.6 & 25.7 & 15.3 & 10.8 &  8.4 & 4.0 & 2.8 & 2.2 & $\lambda=0.05$ \\ \bottomrule
\end{tabular}
\end{table}
Given the steady-state ARL results, we conclude that the EWMA chart easily dominates the DEWMA chart.
From the weighting patterns in \cite{Knot:EtAl:2021a} for the DEWMA chart (and TEWMA) we would expect this steady-state ARL deficiency.
Summing up, the DEWMA (and even more the TEWMA) control chart designs
are not worth the required effort. It is easy to find an EWMA design which
features about the same zero-state ARL values and much better steady-state
ARL results. From \cite{Mahm:Wood:2010} we know that the EWMA chart
exhibits better (lower) worst-case ARL results as well.

\section{DPM Charts} \label{sec:dpm}

\cite{Abba:EtAl:2012} introduced the progressive mean (PM) chart. Afterwards, dozens of derivatives appeared. In \cite{Knot:EtAl:2021a}
it was demonstrated that the PM is not an appropriate control
chart design. However, in \cite{Abba:EtAl:2019a} similarly to the (D)EWMA and (D)MA charts, the double PM (DPM) chart was proposed. Later, in \cite{Riaz:EtAl:2021c}
the variance term used for constructing the alarm rule was corrected.
Following the more recent paper, we describe the DPM design as follows.
\begin{alignat*}{2}
  \text{PM statistic: }\quad && P_t & = \frac{1}{t} \sum_{i=1}^t X_i \,, \\
  \text{DPM statistic:}\quad  && D_t & = \frac{1}{t} \sum_{i=1}^t P_i \,. \\
  \text{PM run-length: }\quad && 
	L_P & = \min \Big\{t\ge 1\!: |P_t - \mu_0| > \big(L_P / t^{0.5} \big) / t^p \,\Big\} \,, \\
  \text{DPM run-length: }\quad && 
    L_D & = \min \Big\{t\ge 1\!: |D_t - \mu_0| > \big(L_D \, \sigma_{D;t} \big)/ t^p \,\Big\} \,.
\end{alignat*}
The exponent $p$ of the curve bending factor $1/t^p$ was set to $p = 0.2$
in \cite{Abba:EtAl:2012}. However, in \cite{Riaz:EtAl:2021c}
$p$ was taken from $\{0.35, 0.40, 0.45, 0.50 \}$ (for PM and DPM).
The variance of $P_t$ is just $1/t$ which explains the $t^{0.5}$ above in
$L_P$. The variance of DPM's $D_t$ is much more complicated.
As mentioned, \cite{Riaz:EtAl:2021c} corrected the old term in 
\cite{Abba:EtAl:2019a} and provided the following term
\begin{equation*}
  \sigma_{D;t}^2 = \frac{1}{t^2} \left\{ \sum_{k=1}^t \left( \sum_{j=k}^t \frac1j \right)^2 \right\}
    = (2t - H_t) / t^2 
\end{equation*}
with the harmonic number $H_t = \sum_{k=1}^t 1/k$.
The above simplification follows from some well-known identities for the partial sums of $H_t$ and $H_t^2$ and, of course, tedious algebra.
By using its approximation
\begin{equation*}
  H_t = \ln t + \gamma + \frac{1}{2t} - \frac{1}{12t^2}
  + \mathcal{O}\left(\frac{1}{t^4}\right)
\end{equation*}
the term used for $\sigma_{D;t}$ is just
\begin{equation*}
  \sigma_{D;t}^2 \approx
  \Big(2t - \ln t -\gamma - 1/(2t) + 1/(12 t^2) \Big) / t^2
\end{equation*}  
with the Euler-Mascheroni constant $\gamma = 0.57721\,56649\,01532 \ldots$ 
Note that this approximation is excellent. However, these mathematical
developments can distract one from the actual problems with the DPM approach.
 
\begin{figure}[hbt]
\centering
\renewcommand{\tabcolsep}{0mm}
\begin{tabular}{cc} 
  \small $\delta = 0.35$ & \small $\delta = 0.75$ \\[-1ex]
  \includegraphics[width=.5\textwidth]{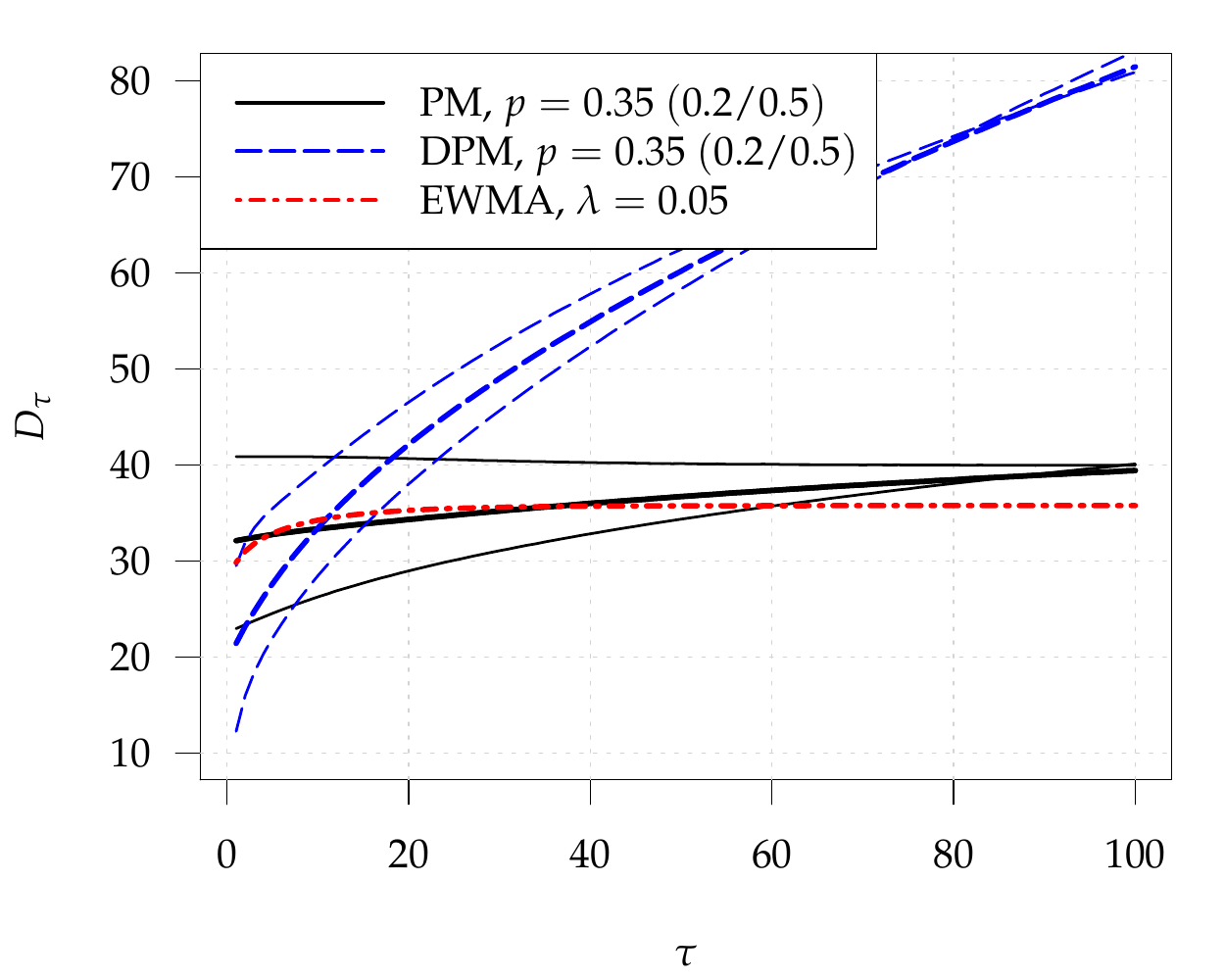} & \includegraphics[width=.5\textwidth]{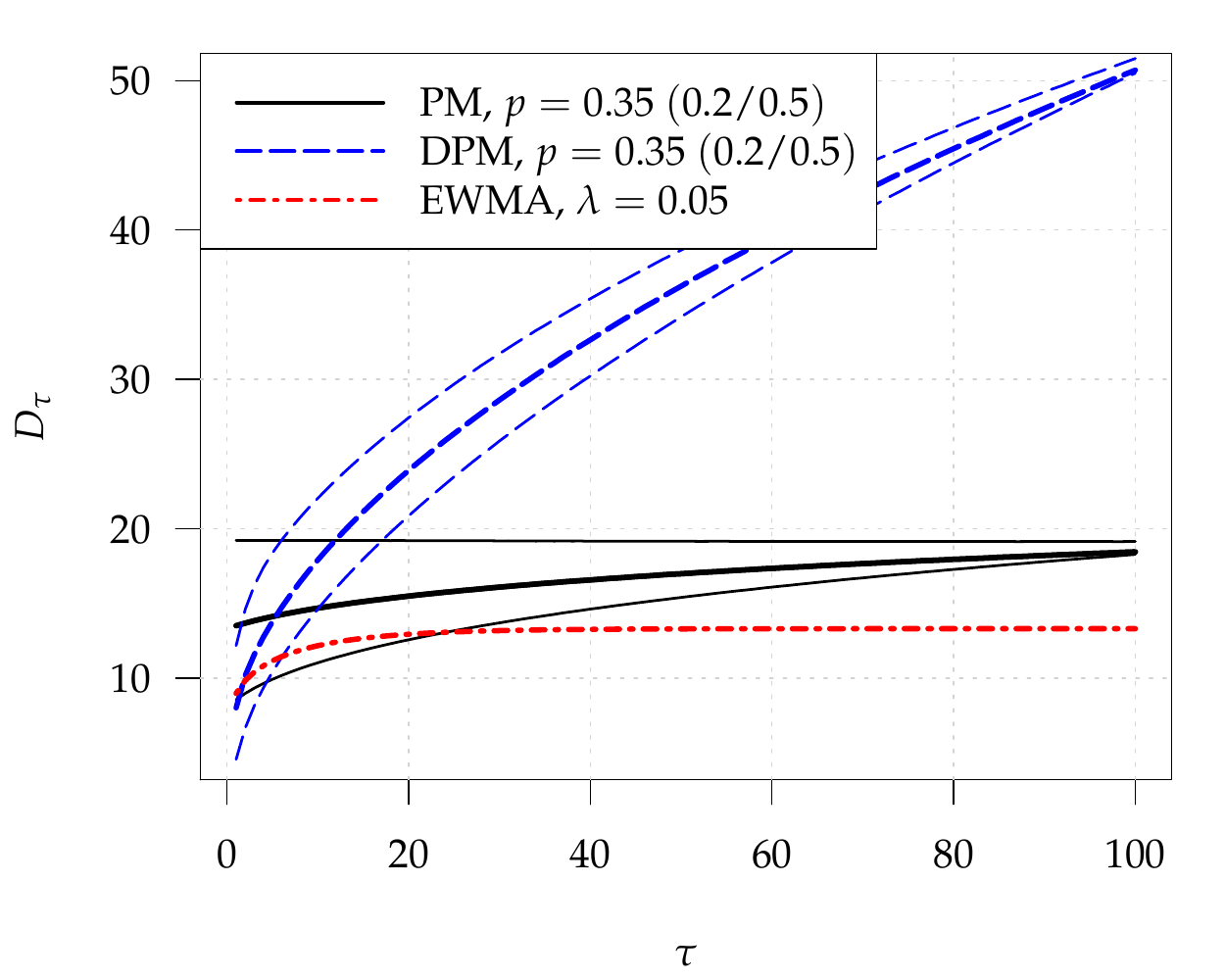} \\[-1ex]
  \small $\delta = 1$ & \small $\delta = 2$ \\[-1ex]
  \includegraphics[width=.5\textwidth]{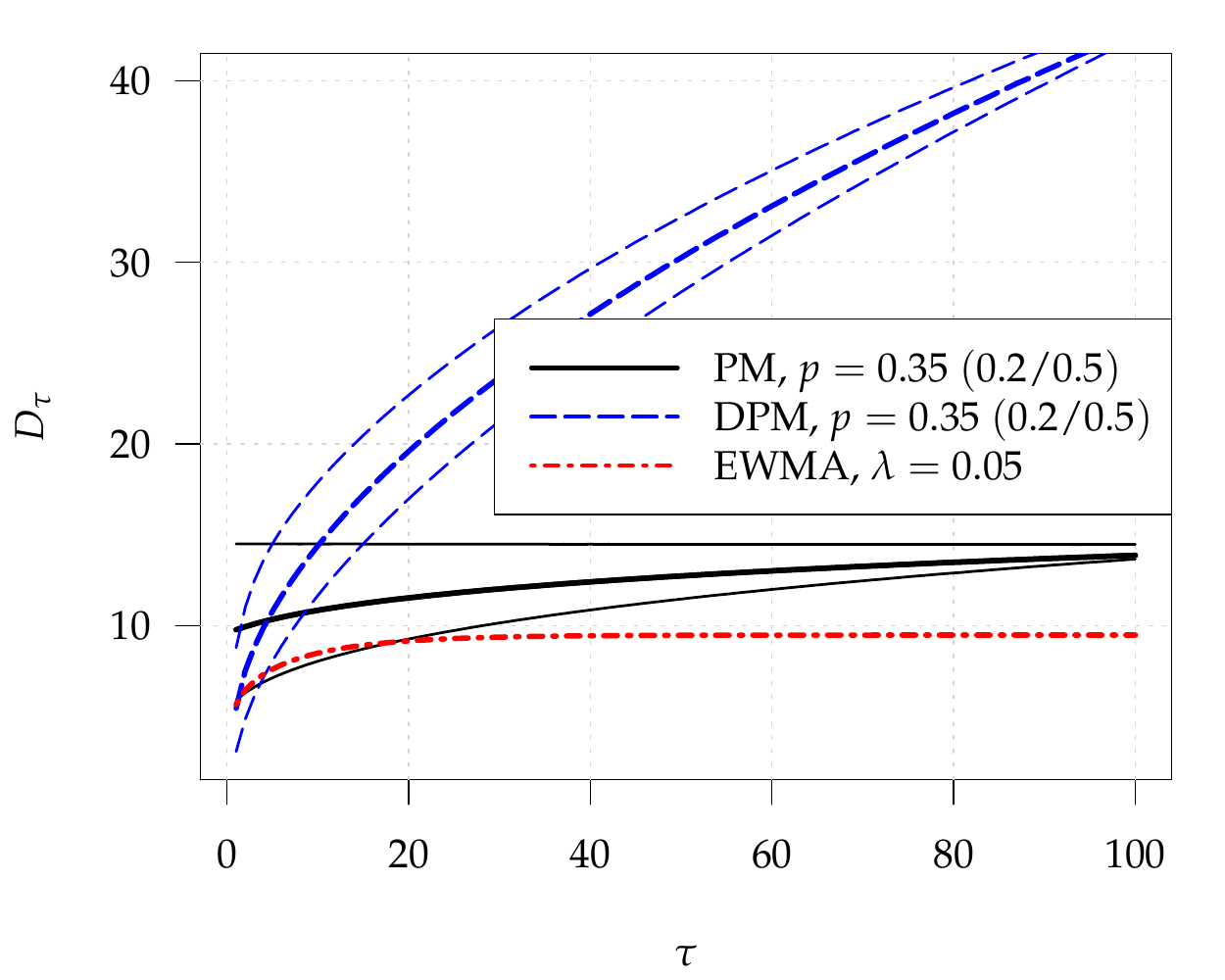} & \includegraphics[width=.5\textwidth]{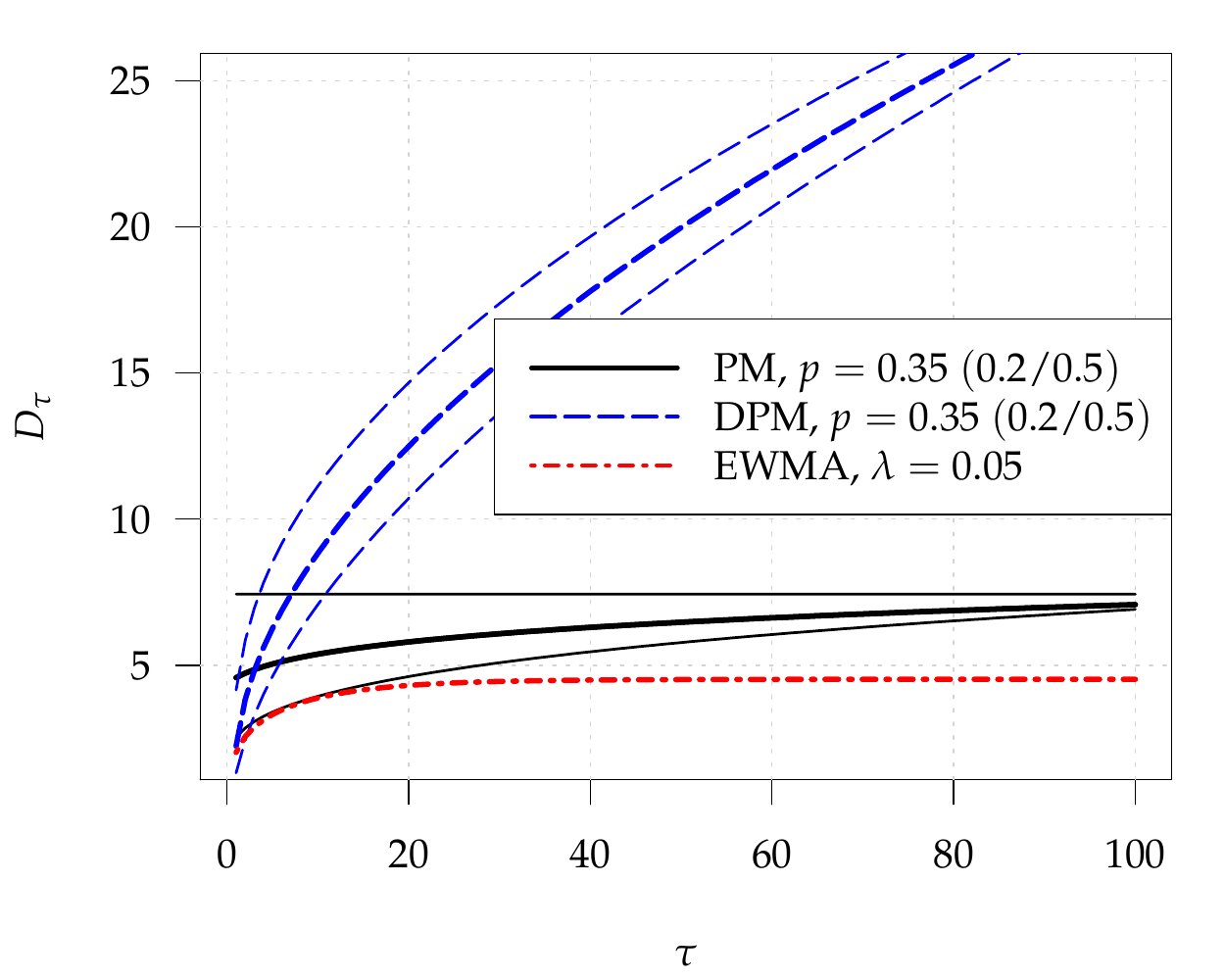}
\end{tabular}
\caption{CED profiles of PM, DPM and EWMA\,($\lambda=0.05$), i.\,e.
$D_\tau = E_\tau(L-\tau+1\mid L\ge \tau)$ vs. change-point position $\tau$, 
for four shifts $\delta$; in-control (zero-state) ARL is $A = 200$.}
\label{fig:dpm_CED}
\end{figure}

For an in-control ARL of 200, we pick $L_P=6.415$ and $L_D=2.596$
from Table 1 and 2 in \cite{Riaz:EtAl:2021c}. ``Our'' competitor is
an EWMA chart with $\lambda=0.05$.
First, we present in Table~\ref{tab:dpm_zARL} some
zero-state ARL results taken from \cite{Riaz:EtAl:2021c}, new ones
with more replications ($10^8$) and EWMA results by deploying the
\textsf{R} package \texttt{spc}.
\begin{table}[hbt]
\centering
\caption{Zero-state ARL numbers from Table 1 and 2 in \cite{Riaz:EtAl:2021c},
$p = 0.35$; EWMA$_1$ with $\lambda = 0.05$ as common design and
EWMA$_2$ ($\lambda=0.007$) as a special one;
in-control ARL 200.}\label{tab:dpm_zARL}
\begin{tabular}{cccccccccc} \toprule
  & \multicolumn{8}{c}{shift $\delta$} \\
  & 0 & 0.25 & 0.35 & 0.5 & 0.75 & 1 & 1.5 & 2 \\ \midrule
  PM   & 198.26 & 47.14 & 32.03 & 21.39 & 13.39 & 9.74 & 6.25 & 4.57 \\
  $10^8$ & 197.86 & 46.80 & 32.09 & 21.40 & 13.52 & 9.78 & 6.24 & 4.58 \\
  DPM  & 201.43 & 31.90 & 21.55 & 13.70 &  8.05 & 5.46 & 3.22 & 2.23 \\
  $10^8$ & 197.50 & 31.65 & 21.40 & 13.69 &  8.02 & 5.46 & 3.20 & 2.23 \\
  EWMA$_1$ & 200.00 & 48.82 & 29.83 & 17.15 &  8.99 & 5.68 & 3.04 & 2.02 \\
  EWMA$_2$ & 200.00 & 30.01 & 18.43 & 10.77 &  5.82 & 3.80 & 2.17 & 1.54 \\
  \bottomrule
\end{tabular}
\end{table}
\cite{Riaz:EtAl:2021c} concluded from a graphical counterpart of our Table~\ref{tab:dpm_zARL} that it \textit{``is obvious from these graphical displays that the DPM chart takes an edge over EWMA, DEWMA, PM, HWMA, and DHWMA chart''}. From Table~\ref{tab:dpm_zARL} we can see, however, that this is an incorrect statement because one could simply decrease the $\lambda$ value to beat the DPM chart
uniformly. However, for PM, DPM and EWMA charts with variance adjusted limits (the one we apply here), we have to consider the CED series and, if applicable, its limit. From the shifts in Table~\ref{tab:dpm_zARL} we consider
$\delta \in \{0.35, 0.75, 1, 2\}$ and change-point positions up to
$\tau = 100$ (in-control ARL 200).
In addition to Table~\ref{tab:dpm_zARL}, we present results as well
in Figure~\ref{fig:dpm_CED} for $p = 0.2$ and $= 0.5$ in order to illustrate the impact of
tuning $p$. Essentially, decreasing $p$ leads to decreased zero-state
ARL, while for late changes the ordering is reversed. The $p = 0.35$
curves of the PM and DPM charts are bold lines.
From all four cases we conclude straightforwardly that for $\tau \ge 30$
the CED $D_\tau$ of the DPM chart is substantially larger than of the PM and EWMA charts.
Even more, the DPM control chart is not able to detect delayed changes ---
it is not too presumptuous to conjecture that DPM's CED series grows in an unbounded way.
Only for $\tau \le 5$ (roughly zero-state) it is usable at all.
In \cite{Knot:EtAl:2021a} the authors showed that the PM chart is not a good choice.
From Figure~\ref{fig:dpm_CED} we can see that the DPM chart is even worse.

\section{Conclusions} \label{sec:conclusions}

We find the lines of research on compound charts outlined in our paper to be misguided. It is very important for any monitoring method, for example, to be based on a weighting scheme that emphasizes the present data values more than the past data values. This rules out use of the DEWMA, triple EWMA, double MA, triple MA, quadruple MA and double PM charts. In addition, PM and HWMA charts should not be used based on their poor steady-state run length performance. The extra complications of the GWMA chart are not justified by any improved performance relative to the simpler EWMA chart. Overall the added complexity of the various proposed compound methods is not justified by improved performance. 

There is an implicit assumption in the literature reviewed here that any shift in the process, however small, is to be detected as quickly as possible. In an increasing number of applications the focus should be on detecting changes of practical importance, not just those resulting in statistical significance. More information can be found in \cite{Wood:Falt:2019}.

\bibliographystyle{unsrtnat}
\bibliography{00_biblio}

\end{document}